\documentclass[twocolumn]{aastex62}

\usepackage{amsmath}
\usepackage{graphicx}
\usepackage{lipsum}
\usepackage{hyperref}
\usepackage{mathtools, nccmath}
\usepackage{longtable}
\usepackage{lineno}

\newcommand{\rev}{}
\newcommand\revv[1]{} 

\newcommand{\nast}{NaSt1}

\graphicspath{{./}{figures/}}

\submitjournal{AAS Journals}

\shorttitle{Discovery of NaSt1 Periodicity}
\shortauthors{Lau et al.}

\begin{document}

\title{Discovery of a \rev{310}-day Period from the Enshrouded Massive System NaSt1 (WR~122)}

\correspondingauthor{Ryan Lau}
\email{ryanlau@ir.isas.jaxa.jp}

\author{Ryan M.\ Lau}
\affil{Institute of Space \& Astronautical Science, Japan Aerospace Exploration Agency, 3-1-1 Yoshinodai, Chuo-ku, Sagamihara, Kanagawa 252-5210, Japan}
\author{Samaporn Tinyanont}
\affiliation{Department of Astronomy and Astrophysics, University of California, 1156 High St., Santa Cruz, CA 95064, USA}
\author{Matthew J. Hankins}
\affil{Department of Physical Sciences, Arkansas Tech University, 1701 N. Boulder Avenue, Russellville, AR 72801, USA}
\author{Michael C.\ B.\ Ashley}
\affiliation{School of Physics, University of New South Wales, Sydney NSW 2052, Australia}
\author{Kishalay De}
\affiliation{Division of Physics, Mathematics, and Astronomy, California Institute of Technology, Pasadena, CA 91125, USA}
\author{Alexei V.\ Filippenko}
\affil{Department of Astronomy, University of California, Berkeley, CA 94720-3411, USA}
\affil{Miller Institute for Basic Research in Science, University of California, Berkeley, CA 94720, USA}
\author{Lynne A.\ Hillenbrand}
\affiliation{Division of Physics, Mathematics, and Astronomy, California Institute of Technology, Pasadena, CA 91125, USA}
\author{Mansi M.\ Kasliwal}
\affiliation{Division of Physics, Mathematics, and Astronomy, California Institute of Technology, Pasadena, CA 91125, USA}
\author{Jon C.~Mauerhan}
\affil{The Aerospace Corporation, 2310 E. El Segundo Boulevard, El Segundo, CA 90245, USA}
\author{Anthony F.~J.~Moffat}
\affil{Département de Physique, Université de Montréal, C.P. 6128, succ. centre-ville, Montréal (Qc) H3C 3J7, Canada; and Centre de Recherche en Astrophysique du Québec, Canada}
\author{Anna M.\ Moore}
\affil{Research School of Astronomy and Astrophysics, Australian National University, Canberra, ACT 2611, Australia}
\author{Nathan Smith}
\affiliation{University of Arizona, Steward Observatory, 933 N. Cherry Avenue, Tucson, AZ 85721, USA}
\author{Jamie Soon}
\affil{Research School of Astronomy and Astrophysics, Australian National University, Canberra, ACT 2611, Australia}
\author{Roberto Soria}
\affil{National Astronomical Observatories, Chinese Academy of Sciences, Beijing 100012, People's Republic of China}
\author{Tony Travouillon}
\affil{Research School of Astronomy and Astrophysics, Australian National University, Canberra, ACT 2611, Australia}
\author{Karel A.\ van der Hucht}
\affil{SRON Netherlands Institute for Space Research Sorbonnelaan 2, NL-3584CA Utrecht, the Netherlands}
\author{Peredur M.~Williams}
\affil{Institute for Astronomy, University of Edinburgh, Royal Observatory, Edinburgh EH9 3HJ, UK}
\author{WeiKang Zheng}
\affil{Department of Astronomy, University of California, Berkeley, CA 94720-3411, USA}




\begin{abstract}

\rev{We present optical and infrared (IR) light curves of NaSt1, also known as Wolf-Rayet (WR) 122, with observations from Palomar Gattini-IR (PGIR), the Zwicky Transient Facility (ZTF), the Katzman Automatic Imaging Telescope (KAIT), the Asteroid Terrestrial-impact Last Alert System (ATLAS), and the All-Sky Automated Survey for Supernovae (ASAS-SN). We identify a $P=309.7\pm0.7$ d photometric period from the optical and IR light curves that reveal periodic, sinusoidal variability between 2014 July and 2021 July. We also present historical IR light curves taken between 1983 July and 1989 May, which show variability consistent with the period of the present-day light curves. In the past, NaSt1 was brighter in the $J$ band with larger variability amplitudes than the present-day PGIR values, suggesting that NaSt1 exhibits variability on longer ($\gtrsim$ decade) timescales. Sinusoidal fits to the recent optical and IR light curves show that the amplitude of NaSt1's variability differs at various wavelengths and also reveal significant phase offsets of $17.0\pm2.5$ d between the ZTF $r$ and PGIR $J$ light curves. We interpret the $310$ d photometric period from NaSt1 as the orbital period of an enshrouded massive binary.  We suggest that the photometric variability of NaSt1 may arise from variations in the line-of-sight optical depth toward circumstellar optical/IR emitting regions throughout its orbit due to colliding-wind dust formation. We speculate that past mass transfer in NaSt1 may have been triggered by Roche-lobe overflow (RLOF) during an eruptive phase of an Ofpe/WN9 star. Lastly, we argue that NaSt1 is no longer undergoing RLOF mass transfer.}

\end{abstract}


\section{Introduction} \label{sec:intro}

Almost all massive stars have a close binary companion ($P\lesssim5000$~d), implying that binary interaction is common in massive stars \citep{Sana2012,Moe2017}.
The ubiquity of binary interactions such as mass transfer has challenged our understanding of massive-star evolution and notably diverges from the theoretical framework of single-star evolution (e.g., \citealt{Heger2003,Groh2013}).
Binary interaction can influence mass loss and enable additional massive-star evolutionary pathways beyond the single-star evolution framework \citep{Smith2014}.
\rev{In particular}, Roche-lobe overflow (RLOF) is a critical binary mass-transfer processes that can strip the hydrogen envelope of a massive star and lead to the emergence of a helium-rich Wolf-Rayet (WR) star \citep{Paczynski1967,Podsiadlowski1992,Demarco2017,Gotberg2018}.
\rev{Agreement between the observed statistical properties of massive stars and predictions from recent binary stellar population synthesis models indeed indicate that binary interactions are important in the formation of WR stars \citep{Vanbeveren1998,Meynet2005,Rosslowe2015,Eldridge2017}.}

The WR phase is an important evolved stage of a massive star that precedes its end-of-life explosion as a Type Ib/c supernova (SN) or direct collapse to a black hole \citep{Heger2003,Smith2011a,Sukhbold2016}. WR binaries may therefore precede the formation of binary black holes and are potential progenitors of black hole mergers (e.g., \citealt{Belczynski2016}).
Populations of WR stars \rev{formed via binary interaction} can also have a profound impact on the ionization of the interstellar medium (ISM) of their host galaxies as well as the intergalactic medium (IGM) beyond \citep{Sander2020,Gotberg2020}. 
WR star dust formation, which has been observed in colliding-winds from carbon-rich WR (WC) stars with OB-star binary companions (e.g., \citealt{Williams1987,Tuthill1999}), may also significantly contribute to dust abundances in the ISM of their host galaxies \citep{Lau2020a}.

Understanding the underlying physics \rev{of binary interactions that lead to the formation of a WR star} is therefore of great importance. However, \rev{there is a dearth of known interacting massive binaries owing to the relatively short timescales associated with the RLOF mass-transfer process in post-main-sequence (Case B) mass transfer.} 
One of the most well-known interacting massive binary systems that is currently undergoing RLOF mass transfer is RY Scuti, which is an eclipsing system with an $\sim$11 day orbital period surrounded by a toroidal circumstellar nebula of dust and ionized gas \citep{Gehrz1995,Grundstrom2007,Smith99,Smith02,Smith2011}. 

The enigmatic \nast~system (also known as LS IV +00 5 and WR~122) was initially classified as a \rev{late-type, nitrogen-rich} WR \rev{(WN10)} star \rev{by \citet{Nassau1963} and \citet{Massey1983} but was later proposed to be a B[e], O[e], or Ofpe/WN9 star \citep{vdH1989,vdH1997}. \nast~again} had its true nature called into question as a result of high spectral resolution observations by \citet{Crowther1999}\rev{, who suggested that \nast~may be an early-type WR star enshrouded by an $\eta$ Car-like nebula}. It is now thought to host a WR star emerging from a RLOF mass-transfer phase with a binary companion \citep{Mauerhan2015}. \nast~\rev{is an intriguing system that may} present a rare opportunity to investigate this transitional phase, particularly given its relatively nearby distance of $3.03^{+0.60}_{-0.45}$~kpc \citep{Rate2020}.

The presence of a binary system in \nast~was inferred from its surrounding extended ($\sim7\arcsec$) \rev{disk-like}, nitrogen-rich nebula that is thought to originate from nonconservative mass transfer in the central system \citep{Mauerhan2015}. \citet{Mauerhan2015} also showed that \nast~exhibits X-ray emission consistent with colliding-wind WR binaries. An important diagnostic of mass transfer via RLOF are the orbital and stellar properties of the central binary; however, these properties remain largely unknown for \nast.  

\nast~possesses an interesting combination of \rev{observational} characteristics that are important for understanding \rev{the nature of its stellar component(s)}. Optical and near-infrared (IR) spectroscopy by \citet{Crowther1999} showed no obvious stellar lines but revealed narrow nebular emission features consistent with fully CNO-processed material that \rev{is thought to be ionized by an unseen WR star.} 
The unusual chemical composition of \nast's nebula has notably drawn comparisons to \rev{the luminous blue variable (LBV) system} $\eta$~Car and its surrounding N-rich nebula \citep{Smith2004}. 
High-excitation emission lines from \nast~also indicated that the ionizing source of its nebula must possess an effective temperature $\gtrsim30,000$~K, which is consistent with the presence of a WR star. Classical WR stars, however, are characterised by broad, full width at half-maximum intensity (FWHM) $\gtrsim1000$~km~s$^{-1}$ emission features\rev{. These broad features} are notably absent from \nast's spectrum. Interestingly, many of the strong emission lines from \nast~exhibit double-peaked profiles separated by tens of km s$^{-1}$, which may originate from a circumstellar or circumbinary disk that could be obscuring or outshining WR emission features \citep{Mauerhan2015}. \rev{Similar multi-peaked narrow emission features dominate the optical spectrum in RY Scuti \citep{Smith02}.} \rev{\nast's nebular spectrum} is also heavily reddened ($A_V\approx6.5$ mag; \citealt{Crowther1999}) \rev{by} interstellar and/or circumstellar material. Strong thermal-IR emission from \nast~indicates the presence of circumstellar dust that either formed recently or is currently \rev{forming} \citep{Rajagopal2007}. 

In this \rev{paper}, we present the discovery of periodic photometric variability from \nast~with data from optical and near-IR imaging surveys. This work also presents some of the first science results from the recently commissioned near-IR imaging survey Palomar Gattini-IR \citep[PGIR;][]{MK2019,De2020a}. \rev{In Section~\ref{sec:2}, we describe the new and archival observations of \nast~used in our analysis. The results of our optical and IR variability analysis of \nast\ are presented in Section~\ref{sec:RA}.  In Section~\ref{sec:discussion}, we discuss the possible origins for the observed photometric variability and speculate on the nature of \nast, including whether it is currently undergoing mass transfer. We summarize our results and conclude in Section~\ref{sec:conclusions}.}

\begin{deluxetable*}{lcccccccccc}
\tablecaption{KAIT, \rev{ATLAS,} ASAS-SN, ZTF, and PGIR Photometry of \nast}
\tablewidth{0pt}
\tablehead{MJD & $B$ & $g$  & $c$ & $V$  & $V$ & $r$  & $R$  & $o$  & $I$  & $J$ \\
 & (KAIT) & (ZTF) & (ATLAS) & (KAIT) & (ASAS-SN) & (ZTF) & (KAIT) & (ATLAS) & (KAIT) & (PGIR)}\
\startdata
56818.446 & 16.21 $\pm$ 0.04 & -- $\pm$ -- & -- $\pm$ -- & 14.38 $\pm$ 0.02 & -- $\pm$ -- & -- $\pm$ -- & 12.82 $\pm$ 0.03 & -- $\pm$ -- & 12.29 $\pm$ 0.03 & -- $\pm$ --\\
56821.406 & 16.23 $\pm$ 0.06 & -- $\pm$ -- & -- $\pm$ -- & 14.39 $\pm$ 0.02 & -- $\pm$ -- & -- $\pm$ -- & 12.82 $\pm$ 0.03 & -- $\pm$ -- & 12.29 $\pm$ 0.03 & -- $\pm$ --\\
56824.434 & 16.22 $\pm$ 0.07 & -- $\pm$ -- & -- $\pm$ -- & 14.38 $\pm$ 0.02 & -- $\pm$ -- & -- $\pm$ -- & 12.81 $\pm$ 0.03 & -- $\pm$ -- & 12.30 $\pm$ 0.03 & -- $\pm$ --\\
56826.437 & 16.18 $\pm$ 0.04 & -- $\pm$ -- & -- $\pm$ -- & 14.38 $\pm$ 0.02 & -- $\pm$ -- & -- $\pm$ -- & 12.80 $\pm$ 0.03 & -- $\pm$ -- & 12.27 $\pm$ 0.03 & -- $\pm$ --\\
... & ... & ...  & ...  & ...  & ...  & ...  & ...  & ... & ...  & ... \\
... & ... & ...  & ...  & ...  & ...  & ...  & ...  & ... & ...  & ... \\
\enddata
\tablecomments{\rev{Optical/IR} photometry from KAIT, \rev{ATLAS,} ASAS-SN, ZTF, and PGIR. All magnitudes are given in the Vega system, and \rev{the errors correspond to the $1\sigma$ uncertainty in the photometric measurements}. A full version of this table (with the KAIT \textit{Clear} photometry) is available electronically.}
\label{tab:LCTabFullShort}
\end{deluxetable*}

\section{Observations and Archival Data}
\label{sec:2}
\subsection{PGIR J-band Photometry}

PGIR is a wide-field, time-domain imaging survey located at Palomar Observatory. It utilizes a 30~cm telescope and $J$-band filter similar to the 2MASS $J$ filter with an effective wavelength of $\lambda_\mathrm{eff}=1.235$ $\mu$m and an effective filter bandwidth of $\Delta \lambda = 0.1624$ $\mu$m. PGIR achieves a $4.96^\circ \times 4.96^\circ$ field of view (FoV) with its 2k $\times$ 2k HAWAII-2RG detector and a scale of $8\farcs7$ pixel$^{-1}$. The entire sky visible from Palomar is imaged by PGIR every $\sim2$ nights down to a 5$\sigma$ detection limit of 14.8 Vega mag\footnote{All magnitudes provided in this work are in the Vega magnitude system.}. The PGIR image processing and photometric calibration are described by \citet{De2020a}. The $J$-band saturation limit of PGIR was initially $\sim8.5$ mag, but has been subsequently improved to $J \approx 6$ mag after implementing a new readout mode described by \citet{De2020b}.

\nast, which is located at $\alpha$(J2000) $= 18^{\rm h}52^{\rm m}17.55^{\rm s}$ and $\delta$(J2000) $= +00^\circ 59' 44\farcs3$ \citep{Gaia18}, was well-covered by PGIR imaging fields. In order to investigate the $J$-band variability of \nast, forced aperture photometry was performed at its coordinates using a 3-resampled-pixel radius ($\sim13\arcsec$) aperture in PGIR images taken between 2018 Nov. and \rev{2021 June.} Importantly, source confusion is not an issue since \nast~is $>3.8$~mag brighter in $J$ than all objects within $15\arcsec$ of its coordinates \citep{2MASS}. 
Photometry from PGIR and all other platforms included in this work are provided in Table~\ref{tab:LCTabFullShort}.

\subsection{KAIT Optical Photometry}
Optical $BVRI$ photometry of \nast~was obtained with the robotic 0.76~m Katzman Automatic Imaging Telescope (KAIT; \citealt{Filippenko2001}) at Lick Observatory. Additional {\it Clear}-band (close to the $R$ band; see \citealt{Li2003}) images were also obtained with KAIT. A series of 66 imaging observations of \nast~was performed with KAIT between 2014 July and 2015 Nov. The KAIT observing frequency of \nast~ranged from a few days to a few weeks when visible.

All images were reduced using a custom pipeline \citep{Ganeshalingam2010,Stahl2019}. Point-spread-function (PSF) photometry was then obtained using DAOPHOT \citep{Stetson1987} from the IDL Astronomy User’s Library\footnote{http://idlastro.gsfc.nasa.gov/}. Several nearby stars were chosen from the Pan-STARRS1\footnote{http://archive.stsci.edu/panstarrs/search.php} catalog for calibration. Their magnitudes were first transformed into Landolt magnitudes \citep{Landolt1992} using the empirical prescription presented by \citet{Tonry2012} and then transformed to the KAIT natural system.

Apparent magnitudes were all measured in the KAIT4 natural system. The final results were transformed to the standard system using local calibrators and color terms for KAIT4 as given in Table 4 of \citet{Ganeshalingam2010}, except for KAIT {\it Clear}-band data, where no reliable color term is measured owing to the broad response function. We therefore present the {\it Clear} magnitude relative to the reference stars in Landolt $R$ magnitude, which is essentially in the KAIT natural system.

\begin{figure*}[t!]
    \centerline{\includegraphics[width=0.99\linewidth]{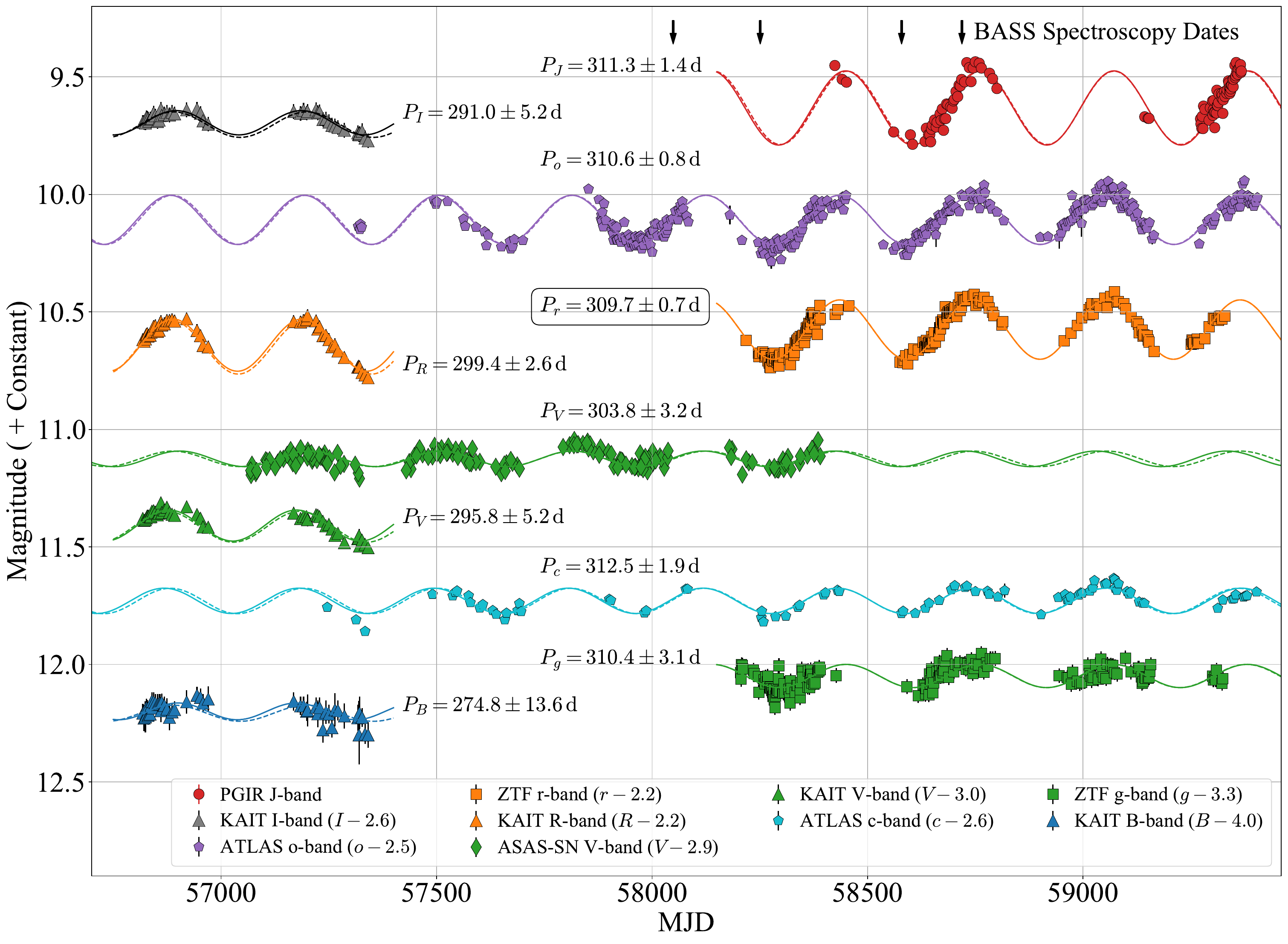}}
    \caption{Optical and near-IR light curves of \nast~from KAIT ($BVRI$), \rev{ATLAS ($co$),} ASAS-SN ($V$), ZTF ($gr$), and PGIR ($J$) taken between 2014 July and \rev{2021 July}. \rev{Constant offsets have been applied to all light curves except that of PGIR, and the offset value is indicated in the legend.} The light-curve plot is overlaid with individual-filter (solid line) and adopted-period (dashed line) sinusoidal fits, where the adopted-period fits \rev{assume the 309.7~d period derived from the ZTF $r$-band fit (shown in box)}. \rev{The similarity of the adopted-period and individual-filter fits of all light curves demonstrate the agreement with a 309.7~d period}. The \rev{periods derived from} individual-filter \rev{fits} are overlaid on the plot next to their respective light curve. Dates of the four mid-IR spectroscopic observations with BASS are also \rev{indicated on the plot as black arrows. Error bars in the light curve correspond to $1\sigma$ uncertainties in the measured magnitudes.}}
    \label{fig:LC}
\end{figure*}

\subsection{BASS Mid-IR Spectroscopy}

Mid-IR observations of \nast~were made using The Broadband Array Spectrograph System (BASS) sensor \citep{Hackwell1990} at the 3.6~m Advanced Electro-Optical System (AEOS) telescope on Haleakela, HI, on 2017 Oct. 23, 2018 May 13, 2019 Apr. 6, and 2019 Aug 24 (UT dates are used throughout this paper).
AEOS utilizes a chopping secondary mirror, run at 7~Hz for rapid sampling of the background. While observing, a chop-and-nod strategy was used with a chop throw of $\sim 11''$ and a telescope nod of the same distance between exposures. 30~s integrations were obtained per nod position, and 12--26 samples were obtained for each of the AEOS observations (6--13~min total integration).

An off-axis visible-wavelength CCD was used for guiding to keep the source centered in the $4.2''$ instrument aperture during exposures. The resolving power of BASS is 30--125 across a 3--13 $\mu$m bandpass.  Flux calibration was obtained via observations of the standard star $\alpha$ Lyr (Vega) at an airmass similar to that of the \nast~observations. \rev{Flux calibration was performed by dividing the target spectrum by the Vega spectrum and then multiplying the result by a Kurucz model atmosphere of Vega matched to the resolution of the data. This process also removes most of the telluric features. However, residual telluric artifacts remained, particularly in the deep water absorption band over 5--8~$\mu$m, so those data were removed for clarity. Although the stellar-atmosphere model does not account for dust excess, Vega's debris excess does not emit significantly shortward of 13~$\mu$m. For example, at the long-wavelength end of our BASS spectrum, Vega's 13~$\mu$m debris disk excess is $< 2$--3\% of the photosphere \citep{Su2013}.} 

Wavelength calibration of each BASS channel was performed in the lab using a monochrometer, and either checked on-sky with observations of a strong mid-IR emission-line source (planetary nebula NGC 7027) or in the dome using a blackbody source (hot plate) measured at two temperatures ($40^\circ$~C and $60^\circ$~C) and attenuated by films of polystyrene and polysulfone to superimpose absorption features for reference.

\begin{deluxetable}{cccccll}
\tablecaption{\nast~Historical IR Magnitudes}
\tablewidth{0.98\linewidth}
\tablehead{MJD  &  $J$ & $H$  & $K$  & $L^{\prime}$ & $M$  & Source}
\startdata
 45528 & 9.54 & 8.51 & 6.50 & 4.12         &            & UKIRT*  \\ 
 45838 & 9.52 & 8.46 & 6.39 & 3.99         & 3.30       & ESO*    \\ 
 46237 & 9.19 & 8.27 & 6.25 & 3.90         & 3.46       & UKIRT  \\ 
 46302 & 9.05 & 7.95 & 5.94 & 3.78         & 3.16       & UKIRT  \\ 
 46715 & 9.49 & 8.18 & 6.19 & 3.91         & 3.18       & UKIRT  \\ 
 46960 & 9.10 & 7.82 & 5.88 & 3.71         & 3.04       & ESO    \\ 
 47258 & 8.99 & 7.85 & 5.87 & 3.66         & 2.98       & ESO \\ 
 47259 & 8.99 & 7.82 & 5.87 & 3.66         & 3.02       & ESO \\
 47261 & 9.04 & 7.89 & 5.87 & 3.68         & 2.96       & ESO \\
 47304 & 9.18 & 7.97 & 6.02 & 3.76         & 3.12       & ESO \\
 47670 & 9.40 & 8.25 & 6.25 & 3.89         & 3.19       & ESO \\
 47671 & 9.39 & 8.25 & 6.25 & 3.90         & 3.23       & ESO \\
\enddata
\tablecomments{IR photometry of \nast~from UKIRT and the ESO 1~m photometric telescope. Uncertainties are assumed to be 0.1~mag for $JHKL'$-band photometry and 0.15~mag for $M$-band photometry.\\ *Previously published by \citep{Williams1987}.}
\label{tab:Hist}
\end{deluxetable}

\subsection{Historical UKIRT and ESO IR Photometry}

We present historical IR $JHKL'M$ photometry of \nast~\rev{(Tab.~\ref{tab:Hist})}~taken from the 3.8~m United Kingdom Infrared Telescope (UKIRT) and the European Southern Observatory (ESO) 1~m photometric telescope between 1983 July and 1989 May (MJD 45528--47671). The UKIRT and ESO photometry taken on 1983 July 13 (MJD 45838) and 1983 May 18 (MJD 45528), respectively, were previously published by \citet{Williams1987}. Subsequent photometry is presented in this work for the first time. Since these observations were taken as a continuation of the photometry published by \citet{Williams1987}, the observations follow the same procedures described in their paper. 
\rev{ESO $JHKL'M$ filter properties are described by \cite{vdB1996}. The UKIRT $JHK$ and $L'M$ filter properties are provided by \cite{Hawarden2001} and \cite{Leggett2003}, respectively.}

UKIRT $JHKL^{\prime}$ photometry was obtained using a 12.4\arcsec-diameter aperture, and $M$ photometry was obtained with a 5\arcsec-diameter aperture. ESO photometry was measured through a 15\arcsec-diameter aperture except for the observation taken on 1987 June 14 (MJD 46960), where a 22\arcsec-diameter aperture was used. The photometry is shown in Table~\ref{tab:Hist} and provided in Vega magnitudes.

\begin{figure}[t!]
    \includegraphics[width=.98\linewidth]{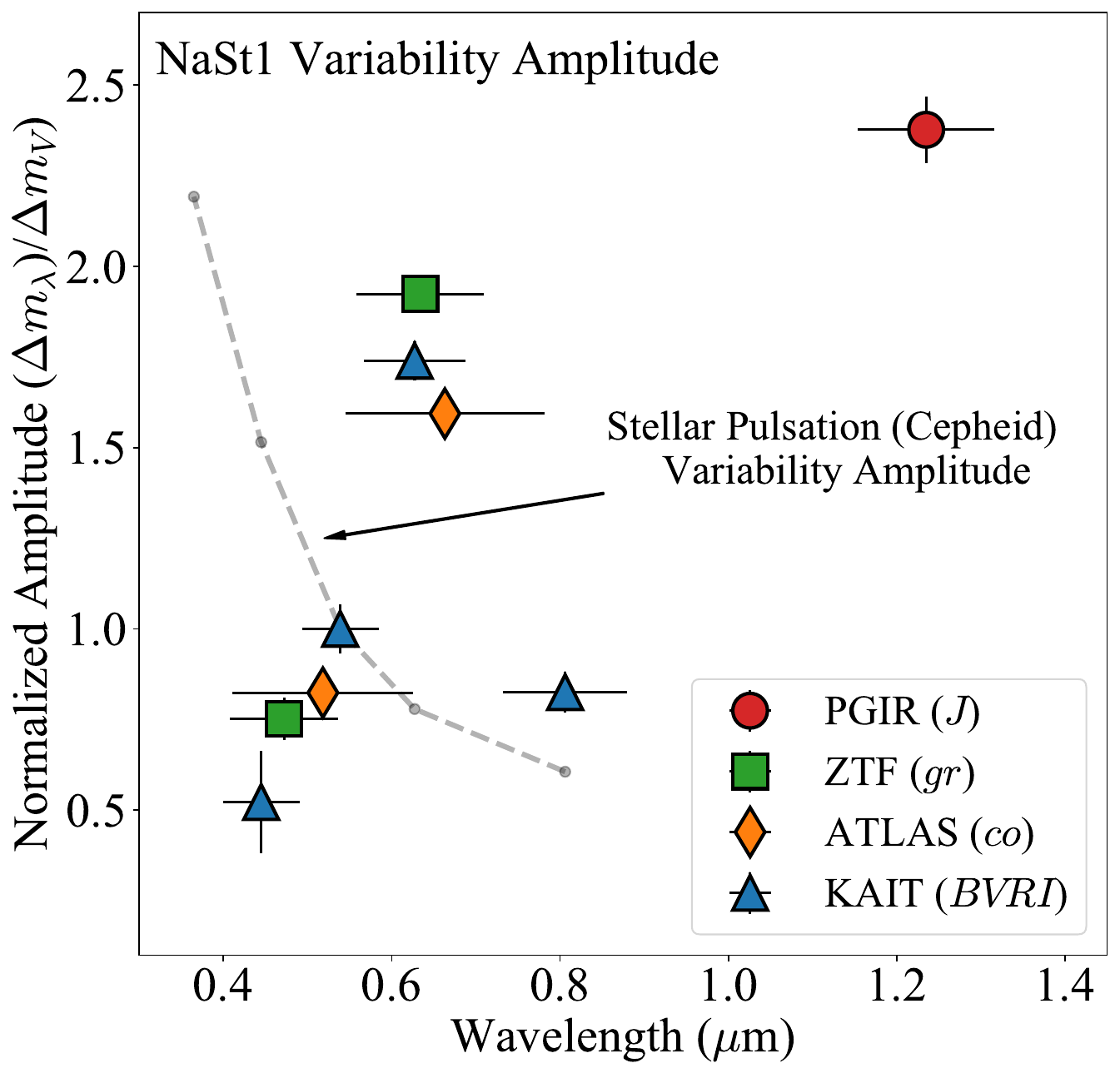}
    \caption{\rev{Normalised} \nast~variability amplitudes from the adopted-period fits to the PGIR, ZTF, \rev{ATLAS,} and KAIT light curves. \rev{The variability amplitudes, $\Delta m_\lambda$,  observed from each filter with effective wavelength $\lambda$ are normalised to the KAIT $V$-band variability amplitude, $\Delta m_V = 0.066\pm0.004$ mag. \rev{Horizontal error bars correspond to the filter bandwidths.} The normalised variability amplitudes observed from stellar pulsations by Cepheids with long periods (log $P>1.02$; \citealt{Klagyivik2009}) are overlaid on the plot.}}
    \label{fig:Amp}
\end{figure}

\begin{figure*}[t!]
    \includegraphics[width=.48\linewidth]{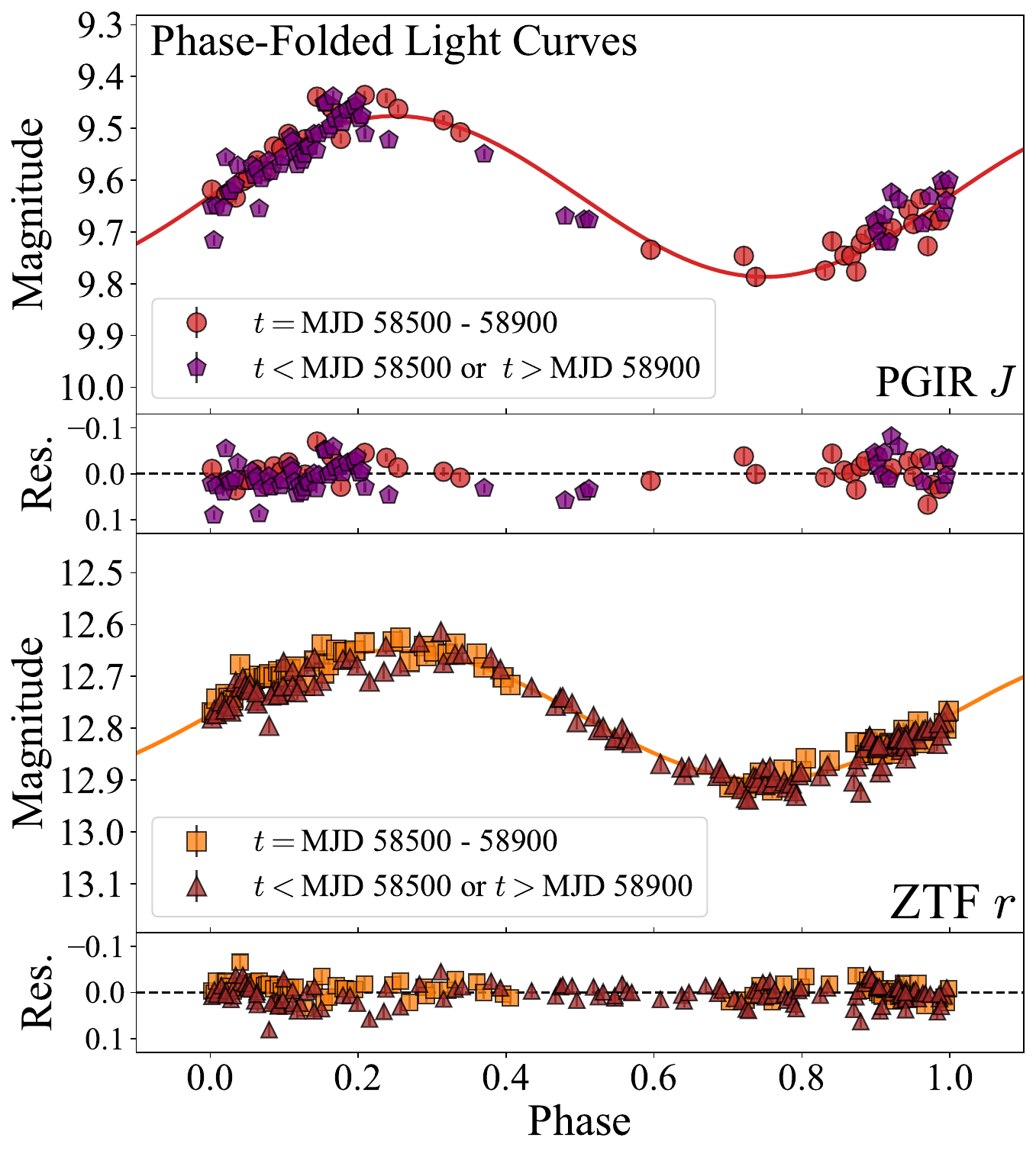}
    \includegraphics[width=.48\linewidth]{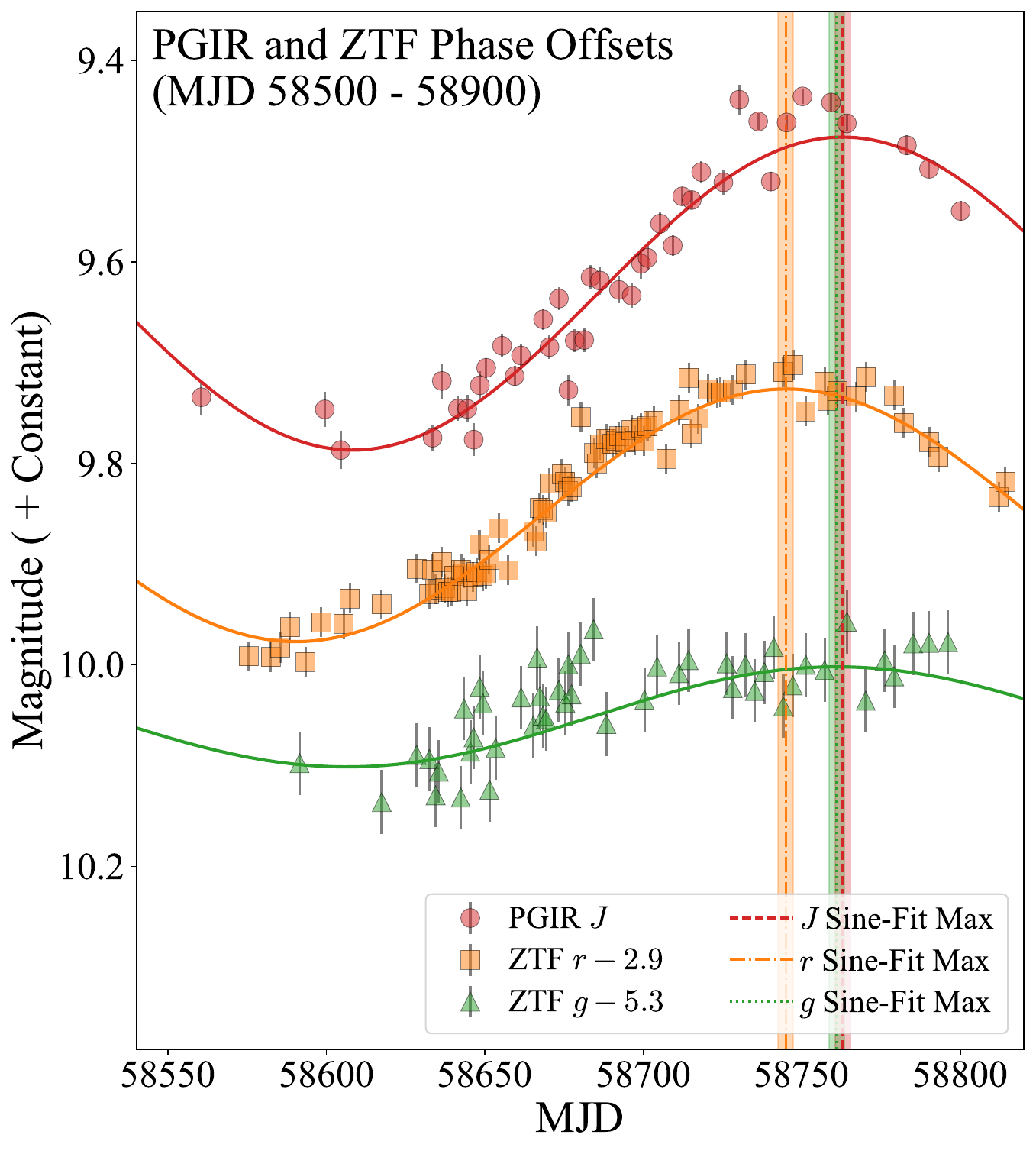}
    \caption{(\textit{Left:}) Phase-folded PGIR $J$-band and ZTF $r$-band photometry overlaid with adopted-period sinusoidal light-curve fits\rev{. Residuals of the light curve are shown below each panel and demonstrate the close agreement with sinusoidal variability.} Observations obtained in the range MJD 58500--58900 are shown as red circles and orange squares for PGIR and ZTF, respectively. Observations obtained before MJD 58500 or after MJD 58900 are shown as purple pentagons or brown triangles for PGIR and ZTF, respectively. (\textit{Right:}) Zoom-in of the PGIR $J$-band and ZTF $r$- \rev{and $g$-}band photometry between MJD 58500 and MJD 58900 overlaid with the adopted-period fits. \rev{Dates of the brightness maxima of each sinusoidal fit are overlaid as vertical lines with the corresponding uncertainties as shaded bands. The offset of the ZTF $r$-band maximum demonstrates the $17.0\pm2.5$ d phase offset from the ZTF $g$- and PGIR $J$-band light-curve fits.}}
    \label{fig:Phase}
\end{figure*}
 
\subsection{Archival Optical Photometry: ZTF, ASAS-SN, \rev{and ATLAS}}
We utilize the $g$-band and $r$-band \rev{PSF-fit} photometry of \nast~from the Zwicky Transient Facility (ZTF; \citealt{Bellm2019,Masci2019}) Public Data Release \rev{6}\footnote{\url{https://www.ztf.caltech.edu/page/dr6}}. ZTF observations of \nast~were taken between 2018 Apr. and \rev{2021 Apr.} with a $\sim1$~d cadence when visible. Only the photometry with clean extractions (i.e., ``catflags = 0'') from ZTF \rev{DR6} were used for the analysis. \rev{Several nights included multiple observations of the field containing \nast. For example,} high-cadence $r$-band observations were taken on MJD 58347 and MJD 58348, where each night consisted of $\sim100$ images. \rev{Multiple ZTF $g$- and $r$-band observations taken over a night were therefore averaged in single-night bins.} Since ZTF photometry from the \rev{DR6} catalog was provided in units of AB magnitude, we converted the photometry to Vega units by applying the conversions $m_{g,\mathrm{AB}} - m_{g,\mathrm{Vega}} = -0.08$ and $m_{r,\mathrm{AB}} - m_{r,\mathrm{Vega}} = 0.16$.

\rev{We also} utilize the $V$-band photometry of \nast~from the All-Sky Automated Survey for Supernovae (ASAS-SN; \citealt{Shappee2014,Kochanek2017}). Photometry of \nast~was obtained from the ASAS-SN Variable Star Database\footnote{\url{https://asas-sn.osu.edu/variables}} \citep{Jayasinghe2020}, where it was identified as a variable with the reference ID ASASSN-V J185217.55+005944.3. ASAS-SN photometry of \nast~\rev{from the Variable Star Database} spans from 2015 Feb. to 2018 Nov. with an observing cadence of $\sim1$ week when visible.

\rev{Lastly, we use photometric measurements of \nast~obtained by the Asteroid Terrestrial-impact Last Alert System (ATLAS) project between between 2015 Oct 23 (MJD 57318) to 2021 Jul 9 (MJD 59404) from the ATLAS forced photometry server\footnote{\url{https://fallingstar-data.com/forcedphot/}} \citep{Tonry2018,Heinze2018,SmithATLAS2020}. ATLAS surveys the entire visible sky at declinations north of $-50^\circ$ every two nights with the ATLAS orange (``o-band"; $\lambda_{\rm eff}=6632$~\AA) and cyan (``c-band"; $\lambda_{\rm eff}=5184$~\AA) filters.
ATLAS photometry obtained from the forced-photometry server were filtered by ignoring observations where the \nast~was within 100 pixels of the edge of the field of view and where the magnitude error was greater than 0.1~mag. 
Multiple observations of \nast~taken over a night were averaged in single-night bins.
PSF photometry of \nast~were converted from Jy to Vega magnitudes using zero points of 3585.79~Jy and 2846.76~Jy for the $c$- and $o$-band filters, respectively, which were derived by the Spanish Virtual Observatory (SVO) Filter Profile Service \citep{Rodrigo2012,Rodrigo2020}. 
}

\begin{deluxetable*}{p{2.5cm}p{2.4cm}p{2.1cm}p{2.1cm}p{2.1cm}cp{1.8cm}}
\tablecaption{\nast~\rev{Adopted-Period Fit Results ($P=309.7$ d)}}
\tablewidth{0.98\linewidth}
\tablehead{Telescope (Filter) & Amplitude (mag) & Mag.~Offset & MJD $(\varphi=0)$ & $\Delta\varphi(r_\mathrm{ZTF})$ (d) & $F_\mathrm{line}/F_\mathrm{phot}$ & Brightest Lines}
\startdata
PGIR ($J$) \newline 12350 \AA~(1624 \AA) &$0.156\pm0.006$ & $9.632\pm0.006$ & $58684.8\pm2.3$ & $17.0\pm2.5$ & $<0.01$ & He\,{\sc i} $\lambda12785$ \newline He\,{\sc i} $\lambda12526$ \newline Pa$\beta$\\
KAIT ($I$) \newline 8061 \AA~(1471 \AA)&$0.054\pm0.004$ & $12.303\pm0.002$ & $56812.0\pm2.7$ & $2.3\pm5.2$ & $0.05\pm0.01$& He\,{\sc ii} $\lambda8237$ \newline O\,{\sc i} $\lambda8446$ \newline [N\,{\sc i}] $\lambda8680$ \\
ATLAS ($o$) \newline 6632 \AA~(2368 \AA)&$0.105\pm0.002$ & $12.608\pm0.002$ & $58667.8\pm1.3$ & $-0.1\pm1.5$ & $0.18\pm0.03$& He\,{\sc i} $\lambda6678$ \newline He\,{\sc i} $\lambda5876$ \newline He\,{\sc i} $\lambda7281$ \\
ZTF ($r$) \newline 6340 \AA~(1515 \AA)&$0.126\pm0.002$ & $12.775\pm0.002$ & $58667.9\pm0.9$ & $0.0\pm0.9$ & $0.23\pm0.03$& He\,{\sc i} $\lambda6678$ \newline He\,{\sc i} $\lambda5876$ \newline H$\alpha$ \\
KAIT ($R$) \newline 6273 \AA~(1202 \AA)&$0.114\pm0.004$ & $12.850\pm0.002$ & $56807.5\pm1.3$ & $-2.1\pm4.6$ & $0.30\pm0.04$& He\,{\sc i} $\lambda6678$ \newline He\,{\sc i} $\lambda5876$ \newline H$\alpha$ \\
ASAS-SN* ($V$) \newline 5466 \AA~(890 \AA)&$0.033\pm0.003$ & $14.025\pm0.002$ & $57120.1\pm4.6$ & $0.8\pm5.9$ & $0.24\pm0.03$& He\,{\sc i} $\lambda5876$ \newline [N\,{\sc ii}] $\lambda5755$ \newline He\,{\sc ii} $\lambda5412$ \\
KAIT ($V$) \newline 5389 \AA~(909 \AA)&$0.066\pm0.004$ & $14.414\pm0.003$ & $56796.2\pm3.2$ & $-13.5\pm5.5$ & $0.12\pm0.02$& [N\,{\sc ii}] $\lambda5755$ \newline He\,{\sc i} $\lambda5016$ \newline He\,{\sc ii} $\lambda5412$ \\
ATLAS ($c$) \newline 5184 \AA~(2144 \AA)&$0.054\pm0.004$ & $14.370\pm0.003$ & $58663.5\pm3.5$ & $-4.4\pm3.6$ & $0.17\pm0.02$& He\,{\sc i} $\lambda5876$ \newline [N\,{\sc ii}] $\lambda5755$ \newline He\,{\sc ii} $\lambda4686$ \\
ZTF ($g$) \newline 4722 \AA~(1282 \AA)&$0.049\pm0.004$ & $15.349\pm0.003$ & $58682.6\pm3.7$ & $14.7\pm3.8$ & $0.15\pm0.02$& He\,{\sc ii} $\lambda4686$ \newline He\,{\sc i} $\lambda5016$ \newline H$\beta$ \\
KAIT ($B$) \newline 4445 \AA~(907 \AA)&$0.034\pm0.009$ & $16.208\pm0.006$ & $56811.9\pm9.4$ & $2.2\pm10.4$ & $0.24\pm0.03$& He\,{\sc ii} $\lambda4686$ \newline H$\beta$ \newline [Fe\,{\sc iii}] $\lambda4658$ \\
\enddata
\tablecomments{\rev{\nast~light curve and adopted-period fit results, which were derived assuming a period matching the ZTF $r$-band value of $P=309.7$~d (see Fig.~\ref{fig:LC}). The listed properties include the telescope/instrument details (filter effective wavelength and bandwidth),} the amplitude of the sinusoidal variability, the magnitude offset at mid-line of the sine function, the MJD where the phase \rev{of the sine function} $\varphi=0$, and the phase offset ($\Delta\varphi$) from the PGIR value. The phase fitting for the ZTF\rev{, ATLAS,} and PGIR models assumed an initial $\varphi=0$ estimate of MJD 58700, while the KAIT and ASAS-SN models assumed initial $\varphi=0$ estimates of MJD 57100 and MJD 56800, respectively. The relative flux contributions from the cataloged nebular emission lines from \nast~\citep{Crowther1999} to the photometric measurements in the different filters wavebands (based on the adopted-period fits) is shown under $F_\mathrm{line}/F_\mathrm{phot}$. The brightest three lines in each filter band is also listed in descending order of line strength. \newline *ASAS-SN photometry of \nast~is likely confused with a nearby ($\sim16\arcsec$) optical point source.}
\label{tab:Prop}
\end{deluxetable*}

\section{Results and Analysis}
\label{sec:RA}

\subsection{Light Curve Periodicity, Variability Amplitude, and Phase Offset}

The near-IR and optical light curves of \nast~taken between 2014 July and \rev{2021 July} are shown in Figure~\ref{fig:LC} and present clear evidence of smooth variability on a roughly year-long timescale. \rev{Since the variations are sufficiently smooth and symmetric about maxima,} sinusoidal functions were fit to the PGIR, ZTF, \rev{ATLAS}, ASAS-SN, and KAIT photometry using a least-squares fitting routine from the \textit{SciPy} library in \textit{Python} v.3.8.5. The free parameters were the period, variability amplitude, magnitude offset, and phase.
The sinusoidal fits to each light curve reveal variability consistent with a period of $\sim300$~d.
\rev{The best-constrained period was derived from the fit to the ZTF $r$-band light curve of $P_r=309.7\pm0.7$~d, which we adopt as \nast's photometric variability period.}
All other light curves were then refitted with an \rev{adopted} period of \rev{$P=309.7$~d.}
The results from \rev{the adopted-period} sinusoidal fits are shown in Table~\ref{tab:Prop}. The following analysis utilizes the results from these \rev{adopted}-period \rev{($P=309.7$~d)} sinusoidal fits.

Interestingly, the sinusoidal variability amplitudes \rev{differ across the different wavelengths}, which is shown in Fig.~\ref{fig:Amp}.
\rev{The ZTF $r$, KAIT $R$, ATLAS $o$, and PGIR $J$ bands exhibit higher variability amplitudes relative to the other photometric bands.}
The KAIT $I$-band amplitude is comparable to the ZTF $g$- \rev{and ATLAS c-band} amplitudes, which \rev{demonstrates that there is no obvious trend in the amplitudes as a function of wavelength}.
\rev{The apparent wavelength dependence of \nast's variability amplitude notably differs from the expected trend for stellar pulsations (e.g.,~Cepheids), where an increased amplitude is expected toward shorter wavelengths \citep{Klagyivik2009}.}

The amplitude fit from the ASAS-SN $V$-band photometry is discrepant with the KAIT $V$-band amplitude despite the similar wavelength coverage of the instrument filters (Tab.~\ref{tab:Prop}). Owing to the much larger $8\arcsec$ pixels of ASAS-SN compared to the $0\farcs8$ KAIT pixels and the two-pixel ($16\arcsec$) radius aperture used for ASAS-SN photometry \citep{Kochanek2017}, the lower ASAS-SN amplitude is likely due to confusion with a nearby source of comparable $V$-band brightness as \nast. Photometry from the Pan-STARRS Data Release 1 catalog \citep{Chambers2016} supports this explanation and shows that there is a source located $\sim16\arcsec$ from \nast~with a PS1 $g$-band brightness $57\%$ that of \nast. 
Synthetic Johnson $V$-band photometry derived from optical spectroscopy of \nast~presented by \citet{Crowther1999}, where $V=14.47$ mag, also shows closer agreement with the 14.41 mag offset derived from the KAIT $V$-band light-curve fit (Table~\ref{tab:Prop}). 
The $0.066\pm0.004$ mag amplitude derived from the KAIT $V$-band photometry therefore more accurately traces the \nast~variability than the lower-amplitude ASAS-SN photometry.

The phase-folded PGIR $J$-band and ZTF $r$-band light curves, which show the highest variability amplitude, are shown in Figure~\ref{fig:Phase} (\textit{left}). \rev{These phase-folded light curves and the fitted-model residuals} demonstrate the close agreement to sinusoidal variability. \rev{The phase-folded light curves also show the  cycle-to-cycle consistency of \nast's photometric variability}.

\begin{figure}[t!]
    \includegraphics[width=.98\linewidth]{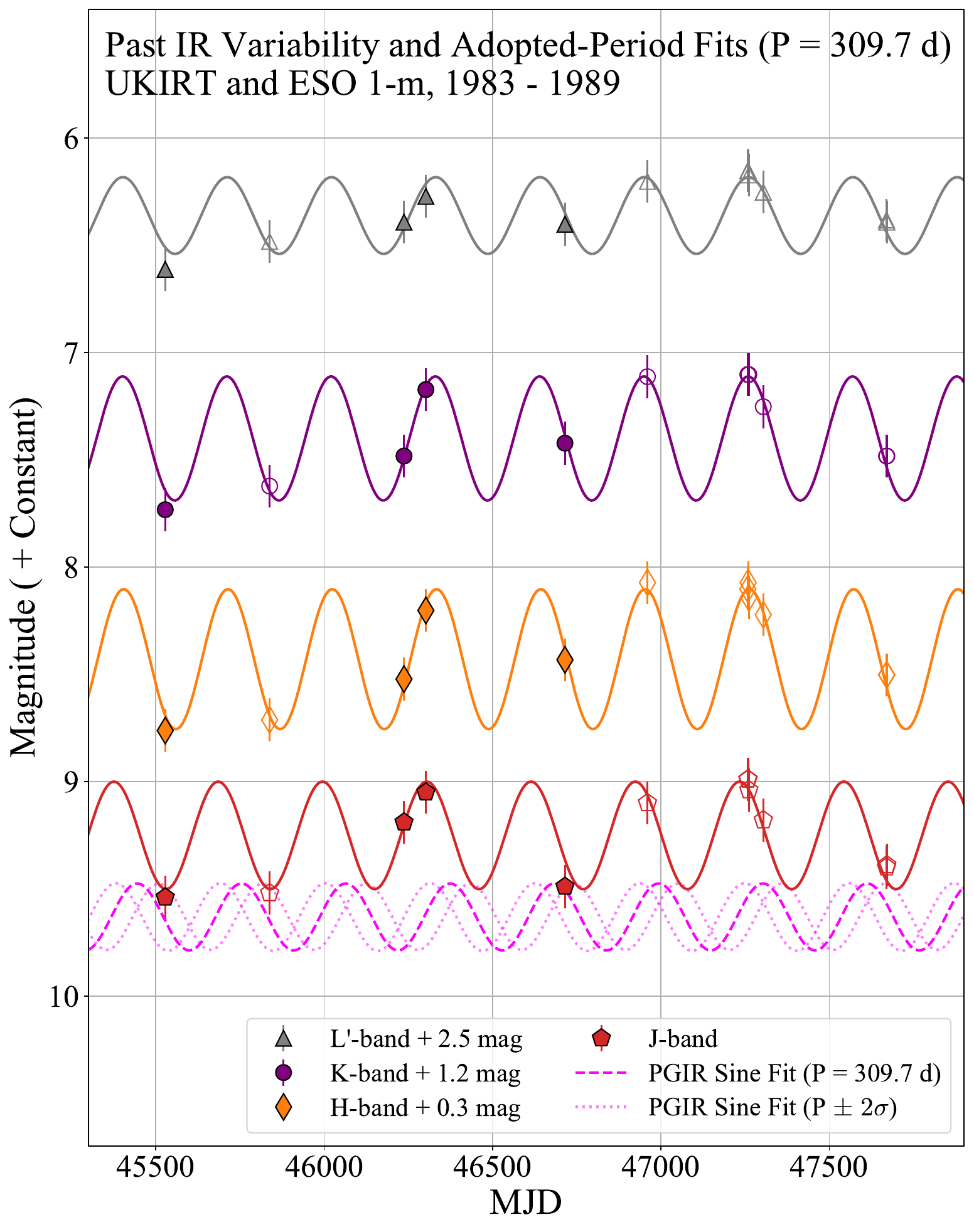}
    \caption{\rev{Adopted-period light-curve fits to} $JHKL'$-band photometry of \nast~observed from UKIRT (filled markers) and the ESO 1~m telescope (open markers) in the years 1983--1989 (MJD 45528--47671). The extrapolated photometric variability from the PGIR $J$-band \rev{adopted}-period light-curve fits \rev{with the $\pm2\sigma$ period uncertainties are plotted (in magenta).}}
    \label{fig:PastVar}
\end{figure}

\begin{deluxetable}{p{2.5cm}p{1.5cm}cc}
\tablecaption{\rev{\nast~Historical IR Light-Curve Fit Results}}
\tablewidth{0.98\linewidth}
\tablehead{Filter  & Amplitude \newline(mag) & Mag.~Offset & MJD ($\varphi=0$) }
\startdata
UKIRT/ESO $J$ \newline U: 1.24~(0.23) $\mu$m\newline E: 1.21~(0.19) $\mu$m & $0.25\pm0.02$ & $9.25\pm0.02$ & $45607.8\pm7.0$\\
UKIRT/ESO $H$ \newline U: 1.65~(0.26) $\mu$m\newline E: 1.64~(0.27) $\mu$m & $0.33\pm0.02$ & $8.18\pm0.02$ & $45636.8\pm3.7$\\
UKIRT/ESO $K$ \newline  U: 2.18~(0.37) $\mu$m\newline E: 2.22~(0.36) $\mu$m & $0.29\pm0.02$ & $6.17\pm0.02$ & $45633.4\pm3.9$\\
UKIRT/ESO $L'$ \newline  U: 3.77~(0.53) $\mu$m\newline E: 3.74~(0.58) $\mu$m & $0.18\pm0.02$ & $3.87\pm0.02$ & $45634.4\pm7.9$\\
\enddata
\tablecomments{\rev{Instrument filter details (effective wavelength and bandwidth) and results from adopted-period ($P = 309.7$~d) sine fits to the historical $JHKL'$-band light curves of \nast~observed by UKIRT and ESO 1~m. Fitted properties include the variability amplitude, magnitude offset at mid-line of the fitted sine function, and the MJD where the phase of the sine function $\varphi=0$. The phase fitting assumed an initial $\varphi=0$ estimate of MJD 45630.}}
\label{tab:HistLCFit}
\end{deluxetable}

The ZTF $r$-band and PGIR $J$-band light curves from \nast\ between MJD 58550 and MJD 58900 and the \rev{adopted}-period sinusoidal fits indicate a significant ($>5\sigma$) phase offset of \rev{$+17\pm2.5$ d, where the $J$-band peak lags behind the $r$-band peak} (Fig.~\ref{fig:Phase}, \textit{right}). 
There is a similar phase offset between the ZTF $r$-band and $g$-band light-curve fits, where the $g$-band offset agrees with the PGIR $J$-band offset within the uncertainties. 
\rev{The phase offsets suggest that the PGIR $J$-band and ZTF $r$-band emission originate from distinct stellar and/or circumstellar components of \nast.}

\begin{figure*}[t!]
    \centerline{\includegraphics[width=0.98\linewidth]{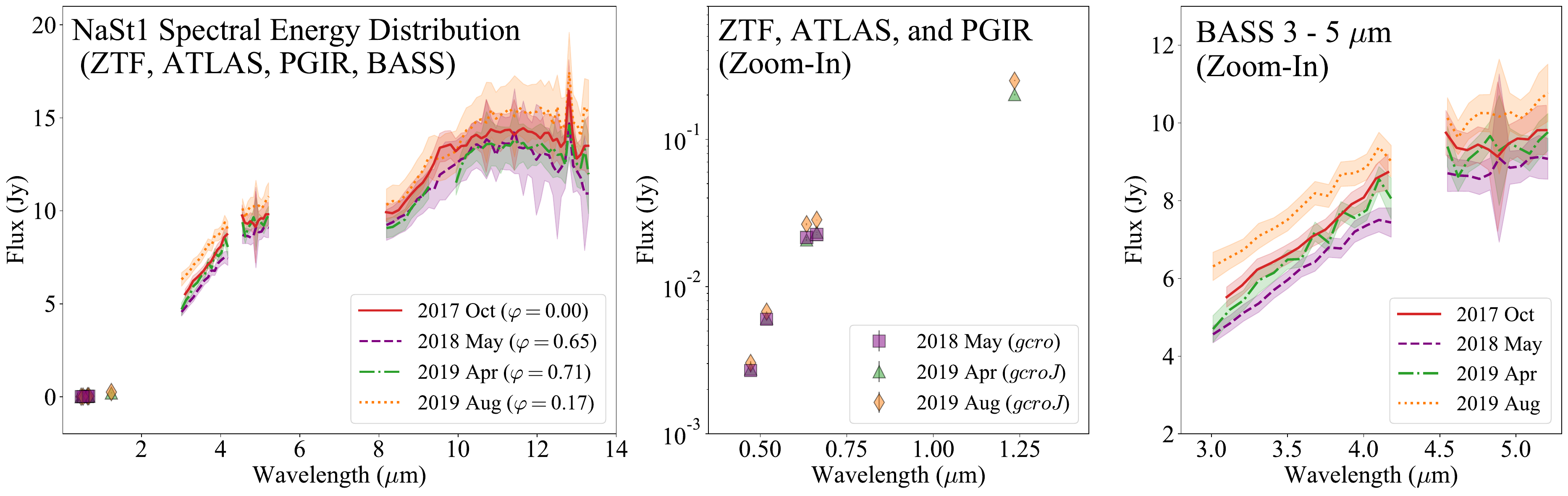}}
    \caption{\rev{(\textit{Left})} \nast~spectral energy distribution variability demonstrated by \rev{4 epochs of} BASS mid-IR 3--13~$\mu$m low-resolution spectroscopy and \rev{3 epochs} of ZTF $gr$-band\rev{, ATLAS $co$-band,} and\rev{/or} PGIR $J$-band \rev{photometric observations}. \rev{BASS observations were taken in 2017 Oct.~(MJD 58049, $\varphi = 0$), 2018 May (MJD 58251, $\varphi = 0.65$), 2019 Apr.~(MJD 58579, $\varphi = 0.71$), and 2019 Aug.~(MJD 58719, $\varphi = 0.17$). Shaded regions correspond to $1\sigma$ flux uncertainties. (\textit{Center}) Zoom-in of the semi-contemporaneous} ZTF\rev{, ATLAS,} and PGIR \rev{observations.} \rev{(\textit{Right}) Zoom-in of 3--5~$\mu$m region of the BASS observations that exhibit the mid-IR variability.} Note that the \rev{zoom-in of the} ZTF\rev{, ATLAS,} and PGIR photometry is plotted on a semilogarithmic scale, whereas the \rev{other plots} are shown on a linear scale.}
    \label{fig:SED}
\end{figure*}

\subsection{Historical IR Variability}
\rev{Adopted-period ($P = 309.7$~d) light-curve fits to the UKIRT and ESO 1~m $JHKL'$-band observations of \nast~taken in 1983--1989 reveal that the periodicity of the past photometric variability is consistent with the present-day observations (Fig.~\ref{fig:PastVar}). \rev{Adopted}-period light-curve fits to the $M$-band observations were unsatisfactory, which was likely due to the larger photometric uncertainties. Fig.~\ref{fig:PastVar} also shows the predicted $J$-band emission based on the \rev{adopted}-period fit to the PGIR light curve. Note that despite the small uncertainties in the period ($1\sigma=0.6$ d) used for the light-curve fitting, the extrapolation from MJD $\sim58600$ to MJD $\sim46500$ leads to large uncertainties in the predicted phase.}
\rev{The consistent fit to the historic photometry strengthens our adopted period and demonstrates coherence of the periodic variability over decades.}

\rev{The results from the adopted-period fits to the historical IR light curves are shown in Table~\ref{tab:HistLCFit}. Similar to the results from the recent optical/IR light curve analysis (Tab.~\ref{tab:Prop}), the variability amplitudes differ at different filter wavelengths. Interestingly, the results from the light-curve fits indicate that the past $J$-band emission was brighter and showed variability amplitudes larger than the recent PGIR $J$-band observations.}

\rev{It is important to consider the color and magnitude conversions between the different IR photometric systems of UKIRT/ESO and PGIR. Based on the UKIRT and 2MASS transformation equations~37--40 from \citet{Carpenter2001} and the measured magnitude offsets from the adopted-period light-curve fits (Tab.~\ref{tab:HistLCFit}), the 2MASS and UKIRT $J$-band photometry conversion is $J_\mathrm{UKIRT}=J_\mathrm{2MASS}-(0.21\pm0.06\,\mathrm{mag})$. Owing to their similar filter properties, $J_\mathrm{2MASS}\approx J_\mathrm{PGIR}$, which implies that the change in $J_\mathrm{PGIR}$ between the past and recent observations is $\Delta J_\mathrm{PGIR}=0.17\pm0.06$ mag. This suggests that \nast~does indeed exhibit variability on longer ($\gtrsim$ decade) timescales.
}

\subsection{Mid-IR Variability}

Mid-IR spectroscopy of \nast~with BASS taken in 2017--2019 reveals variability at 3--13 $\mu$m (Fig.~\ref{fig:SED}), where the mid-IR flux from \nast~is likely dominated by thermal emission from circumstellar dust \citep{Crowther1999,Rajagopal2007}. The IR-dominated spectral energy distribution (SED) in Figure~\ref{fig:SED} \rev{(\textit{left})} also demonstrates that \nast~is enshrouded by circumstellar dust in addition to being obscured by interstellar material.

\rev{Figure~\ref{fig:SED} (\textit{center}) shows the} semicontemporaneous ZTF $gr$\rev{, ATLAS $co$,} and PGIR $J$ photometry taken within 20~d ($\Delta\varphi \approx 0.06$) of the BASS spectroscopy in 2019 Apr. (MJD 58579) and within \rev{7} days ($\Delta\varphi \approx 0.02$) of BASS spectroscopy in 2019 Aug. (MJD 58719). 
\rev{Figure~\ref{fig:SED} (\textit{center}) also shows the ZTF and ATLAS photometry taken within 4~d ($\Delta\varphi \approx 0.01$) of the BASS spectroscopy in 2018 May (MJD 58251).}
\rev{The timing of the 2019 Apr.~($\varphi = 0.71$) and 2019 Aug.~($\varphi = 0.17$) observations sample} near the \nast~minimum and peak emission, respectively (\rev{see} Fig.~\ref{fig:LC}).
The \nast~SED at these two epochs shows that the trend in the mid-IR variability is consistent with the optical/near-IR variability (Fig.~\ref{fig:SED}\rev{, \textit{right}}). 
\rev{The ZTF, ATLAS, and PGIR photometry increased by 10--30\%} between the 2019 Apr.and 2019 Aug.
The 3--13 $\mu$m emission increased \rev{similarly by} $\sim10$--30\% between 2019 Apr. and 2019 Aug., where the largest variations ($>20\%$) were observed at shorter mid-IR wavelengths ($\lambda \approx 3$ $\mu$m).
Variability from the circumstellar dust emission in the mid-IR is therefore likely linked to the optical/near-IR variability.

\subsection{Color Variability}
\begin{figure*}[t!]
    \includegraphics[width=.487\linewidth]{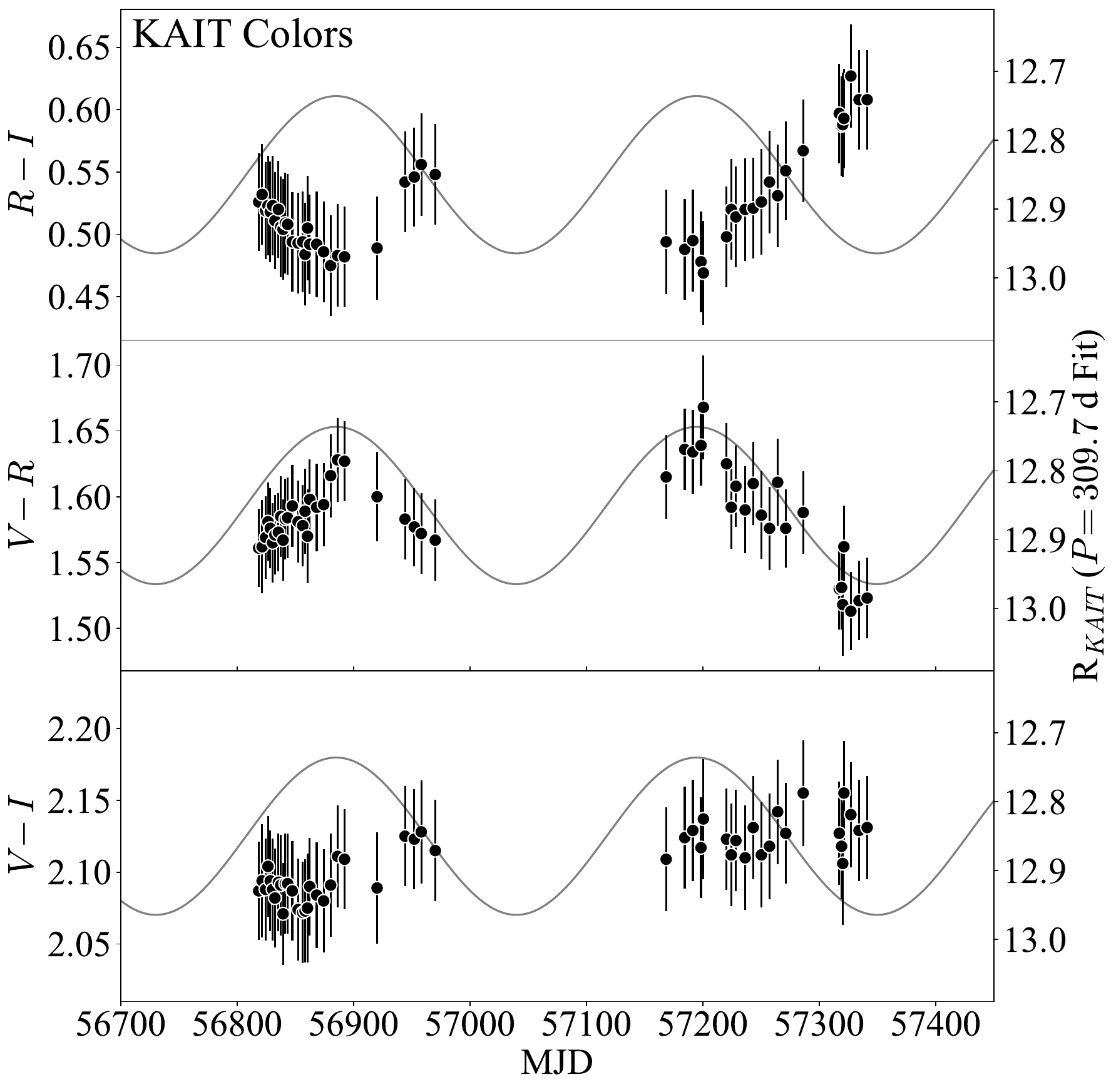}
    \includegraphics[width=.48\linewidth]{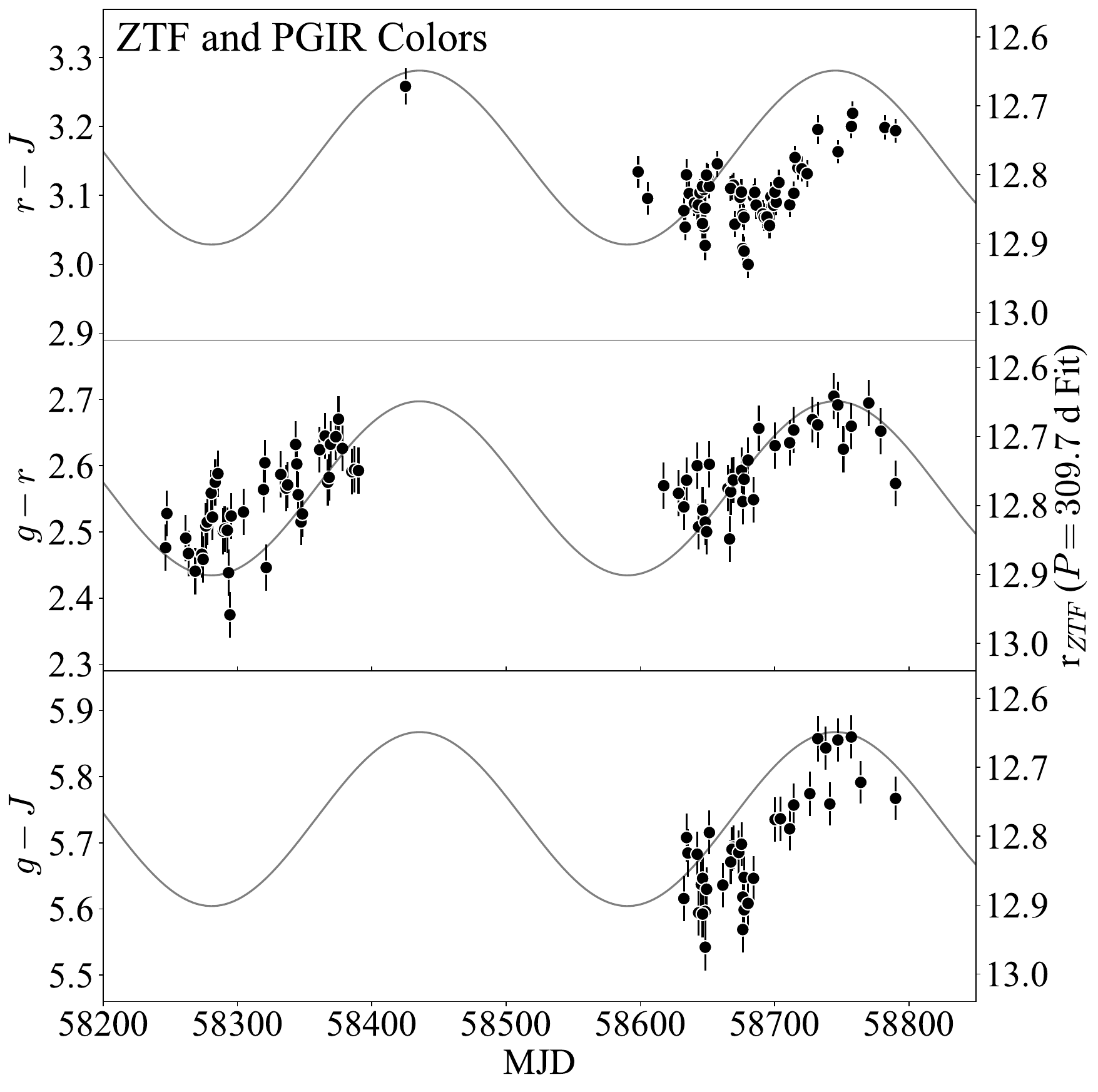}
    \caption{(\textit{Left}) KAIT $V-I$, $V-R$, and $R-I$ colors between MJD 56800 and MJD 57350 overlaid with the KAIT $R$-band \rev{adopted}-period sinusoidal model light curve. (\textit{Right}) ZTF and PGIR $g-J$, $g-r$, and $r-J$ colors between MJD 58200 and MJD 58800 overlaid with the ZTF $r$-band light curve \rev{fit}.}
    \label{fig:Color}
\end{figure*}

In order to investigate the color variability of \nast, KAIT and ZTF/PGIR color curves were produced with simultaneous or nearly contemporaneous observations. Figure~\ref{fig:Color} (\textit{left}) presents KAIT $V-I$, $V-R$, and $R-I$ color curves, which are derived directly from the simultaneous multiband KAIT observations. Since PGIR and ZTF observations were not performed simultaneously, the PGIR/ZTF $g-J$, $g-r$, and $r-J$ light curves in Figure~\ref{fig:Color} (\textit{right}) were produced by utilizing observations taken within $<3$~d ($\Delta\varphi < 0.01$). 

KAIT color curves in Figure~\ref{fig:Color} (\textit{left}) each show different variability trends. The $V-R$ \rev{color} curve \rev{varied consistently with the} $R$-band brightness; however, the $R-I$ curve indicates the opposite trend. The $V-I$ \rev{color curve exhibits} no significant variability\rev{, which is}
likely due to the lower amplitude of the $I$-band variability relative to that of the shorter wavelength $R$-band filter (Fig.~\ref{fig:Amp}).
The ZTF and PGIR $g-J$, $g-r$, and $r-J$ color curves all \rev{show variability} correlated with increasing $r$-band brightness. The discrepancies in the color evolution reflect the inconsistency of the variability amplitude from \nast~as a function of wavelength \rev{and suggest that different wavelength filters trace emission from different stellar and/or circumstellar components}. The observed phase offsets between different filters (Fig.~\ref{fig:Phase}, \textit{right}) likely influence the color evolution as well.

\subsection{Emission-Line Contributions}
\label{sec:Eline}

Optical and near-IR spectroscopy of \nast~by \citet{Crowther1999} revealed prominent narrow nebular emission lines such as He {\sc i} $\lambda 6678$, H$\alpha$, and [N {\sc ii}] $\lambda5755$, which may contribute to the photometric measurements \rev{and variability}. In order to determine whether the emission lines are influencing the variability, the \nast~emission-line list from \citet{Crowther1999} was used to quantify the relative observed flux density contribution to the model-\rev{derived} photometry (i.e., ``Mag.~Offset" in Table~\ref{tab:Prop}). 
A $10\%$ absolute flux calibration uncertainty was assumed for both the emission-line measurements from \citet{Crowther1999} and the PGIR, KAIT, \rev{ATLAS,} ZTF, and ASAS-SN photometry.

\begin{figure*}[t!]
    \centerline{\includegraphics[width=0.96\linewidth]{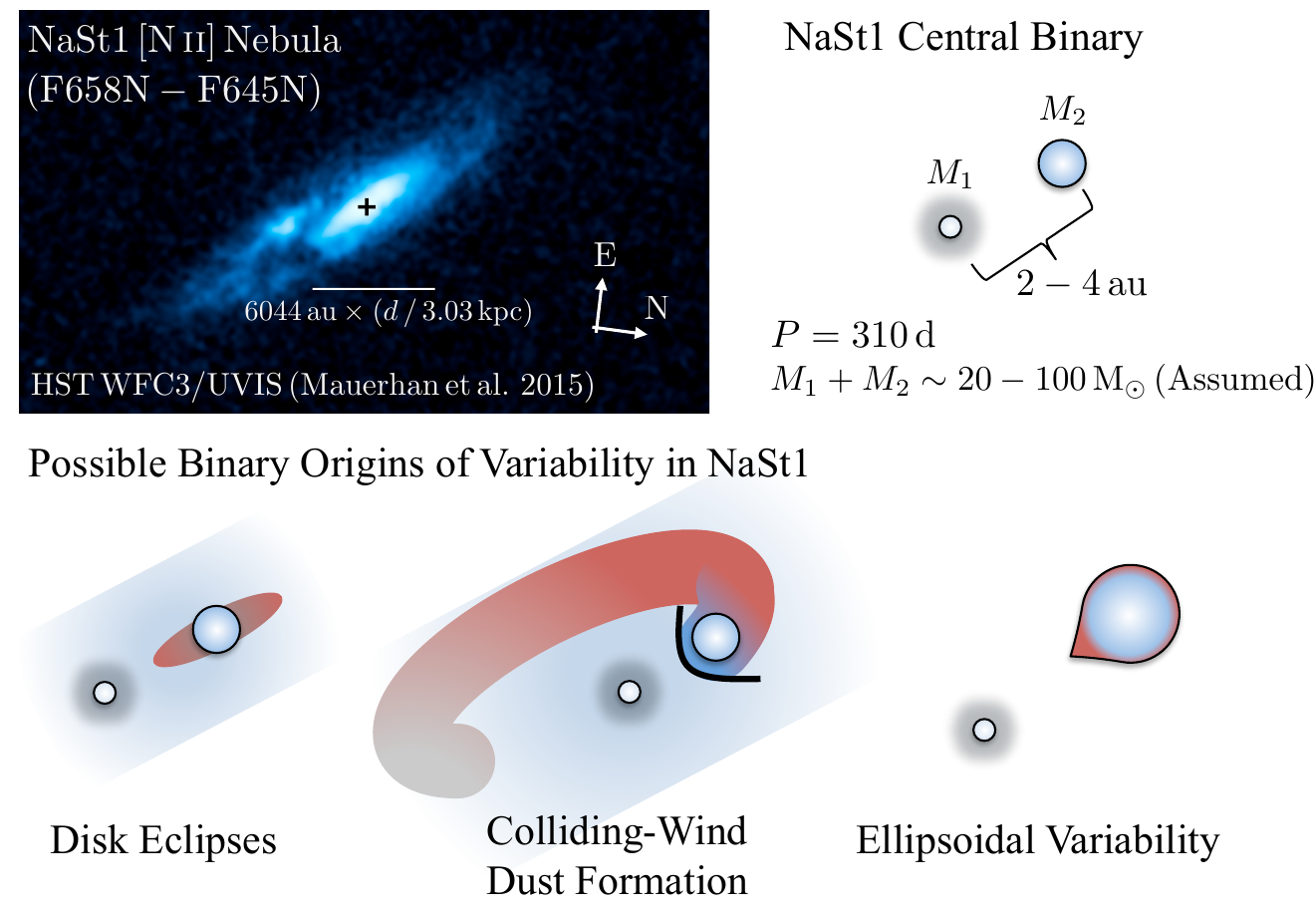}}
    \caption{(Top left:) HST/WFC3/UVIS image of the extended [N\,{\sc ii}] nebula around \nast~from \citet{Mauerhan2015} and (Top right:) schematic illustration of the \rev{possible} binary \rev{in} \nast~overlaid with the \rev{estimated} orbital properties \rev{assuming the $310$~d variability period corresponds to its orbital period in a circular orbit.} $M_1$ \rev{would be} the \rev{mass donor} star and $M_2$ \rev{would be} the mass gainer. (Bottom:) Schematic illustrations of the possible \rev{binary} origins of variability in \nast: disk eclipses, colliding-wind dust formation, and ellipsoidal variability. Note that the illustrations in this figure are not drawn to scale.}
    \label{fig:Schem}
\end{figure*}

Table~\ref{tab:Prop} shows $F_\mathrm{line}/F_\mathrm{phot}$, the ratio of the flux density from \nast's observed emission lines over the filter bandwidths relative to the observed photometry. The nebular emission lines from \nast\ do indeed appear to contribute \rev{$\sim5$--30\%} to the \rev{optical} photometry, 
where the strongest emission lines in \nast~are associated with He\,{\sc i}, He\,{\sc ii}, and [N\,{\sc ii}]. However, the emission line contribution is much lower in $J$. Since the amplitudes in the \rev{optical} light curves are in the range $\sim0.03$--0.13 mag, or $\sim \pm 3$--11\% of the ``Mag.~Offset," it is possible that the photometric variability may be due to changes in the observed strength of the emission lines in these filters. 
\rev{Interestingly, the ATLAS $o$, ZTF $r$, and KAIT $R$ observations which exhibit the largest photometric variability amplitudes at optical wavelengths  (Fig.~\ref{fig:Amp}) are the only bands that include coverage of He\,{\sc i} $\lambda6678$, the strongest optical emission line observed from \nast~ \citep{Crowther1999}. These are also the only filter bands that include coverage of H$\alpha$ and [N\,{\sc ii}] $\lambda6583$, where emission from the latter appears as an extended and asymmetric nebula around \nast\ (\citealt{Mauerhan2015}; Fig.~\ref{fig:Schem}, \textit{top})}

\rev{The emission-line contribution to the PGIR $J$-band photometry is negligible, which indicates that the large $J$ variability amplitude does not arise from variable near-IR emission lines.}
\rev{There is also minimal ($F_\mathrm{line}/F_\mathrm{phot}<0.01$) emission-line contribution to the past $JHK$ photometry from UKIRT/ESO (Tab.~\ref{tab:HistLCFit}), which suggests that the apparent change in the $J$-band amplitude variability between the UKIRT/ESO and PGIR observations is not due to emission line variations.
$J$-band emission from \nast~may instead be dominated by thermal emission from circumstellar dust.
}

\rev{It is possible that the PGIR $J$, ATLAS $o$, ZTF $r$, and KAIT $R$ bands exhibit high-amplitude variability because the filter bands trace emission with a significant contribution from circumstellar material. The other filter bands with lower variability amplitudes may instead be more influenced by stellar continuum emission.}

\section{Discussion}
\label{sec:discussion}

\subsection{Possible \rev{Single-Star} Origins of Variability}
\label{sec:VarOrg}

\rev{\nast's asymmetric nebula (Fig.~\ref{fig:Schem}, \textit{top}; \citealt{Mauerhan2015}) combined with the discovery of its regular period suggests that \nast~hosts a binary. Photometric variability arising from pulsations or other intrinsic instability in hot, single, massive stars also do not typically appear as regular or sinusoidal as we observe from \nast~(e.g., \citealt{Soraisam2020}). We therefore postulate that the smooth $310$ d photometric variability is most likely modulated by the orbit of a binary in \nast. }
\rev{This discussion is notably reminiscent of the long debate on the binary vs. single-star hypotheses on $\eta$ Car, where the binary interpretation is favored based on observations of its stable periodic variability \citep{Damineli1996,Corcoran2005,Damineli2008}.}
\rev{However, it is still important to consider the possible single-star mechanisms that may also account for the observed variability in \nast.}

\rev{One of the most common origins of stellar variability is stellar pulsations. This mechanism can result in near-sinusoidal photometric variations, which have been observed from stars such as Cepheids, Miras, and red supergiants (RSGs). However, the amplitude of the photometric variability arising from stellar pulsations exhibits a characteristic wavelength dependence, where the amplitudes increase toward shorter wavelengths owing to associated changes in the star's effective temperature (e.g.,~\citealt{Klagyivik2009}). This is inconsistent with the observed variability amplitudes of \nast\ at different-wavelength filters as shown in Fig.~\ref{fig:Amp}. We therefore argue against stellar pulsations as the origin of \nast's photometric variability.}

\rev{Another possibility for a single-star variability origin is rotational modulation that occurs in a rare class of highly magnetised O-stars with the peculiar ``Of?p'' classification \citep{Walborn1973,Howarth2007, Grunhut2017}. Of?p stars exhibit near-sinusoidal variability on the order of $\sim0.05$ mag and are attributed to phase-dependent occulations of a wind-trapped magnetosphere \citep{Naze2015, Munoz2020}. They can also be slow rotators with rotation periods that range from weeks to years \citep{Naze2015}. Of?p stars also exhibit dramatic spectral variability. For example, the well-known Galactic Of?p star HD~191612 exhibits both photometric variations and strongly variable Balmer and He\,{\sc i} lines on a $540$ d period \citep{Howarth2007}. Interestingly, another Galactic Of?p star, HD~148937, is surrounded by an extended dusty and N-rich circumstellar nebula \citep{Mahy2017}, which is similar to the properties of \nast's nebula. The photometric variability of Of?p stars, however, is modulated by wavelength-independent electron scattering opacity from a wind-trapped magnetosphere \citep{Munoz2020}. The variability amplitudes should therefore show no significant deviations as a function of wavelength. \citet{Howarth2007} also show that the photometric variations in HD~191612 must arise from continuum-level variability rather than changes in the line emission strength. Although it is unlikely, we cannot completely rule out the Of?p phenomenon as a contributor to \nast's photometric variability.
}

\subsection{Possible \rev{Binary} Origins of Variability}
\label{sec:VarOrgB}

\rev{Given our discussion in the previous section, we assume that \nast's photometric variability} is associated with the orbital period of a binary system as opposed to a single-star origin. \rev{Here, we examine} the possible \rev{binary} origins of \nast's variability and the relation to its orbital period. 
\rev{First, we can infer some orbital properties of the binary system assuming that the orbital period is consistent with the observed $310$~d variability period.}
Assuming a combined stellar mass in the range $\sim20$--100~M$_\odot$ for the binary and a circular orbit, \rev{a $310$~d} orbital period corresponds to an orbital separation of $\sim2$--4 au. We note that a circular orbit is expected for a system that recently underwent RLOF. The orientation of the binary system can be inferred from \rev{\nast's} extended N-rich nebula, which exhibits a nearly edge-on geometry with an inclination angle of $i=78^\circ \pm 3^\circ$ \citep{Mauerhan2015}. Figure~\ref{fig:Schem} (top) shows the resolved {\it HST} imaging by \citet{Mauerhan2015} of \nast's N-rich nebula and our schematic illustration of the embedded central binary. 
\revv{In this section, we discuss three possible binary-related scenarios given \nast’s 310-d period and smooth sinusoidal
variations}
(Fig.~\ref{fig:Schem}, bottom): disk eclipses, colliding-wind dust formation, and ellipsoidal variability.
\revv{We acknowledge that other scenarios are also possible given the limited observational constraints, but for the sake of brevity they will not be considered here.}
\subsubsection{Disk Eclipses}
Given \rev{a} large orbital separation ranging over several au, it is highly unlikely that the smooth, sinusoidal variability \rev{could arise from} eclipses from the stars themselves, which would instead appear as narrow eclipses. However, we suggest that the variability may be caused by eclipses of a circumbinary disk from an optically and geometrically thick circumstellar disk in \nast~that formed as a result of the RLOF mass-transfer process.
Several prominent emission lines in \nast~(e.g., [N\,{\sc ii}] $\lambda5755$; \citealt{Crowther1999}) show double-peaked profiles that indeed indicate the presence of a disk.

This scenario is similar to the massive eclipsing binary RY Scuti, which is currently undergoing RLOF with a mass gainer that appears to be enshrouded by an optically and geometrically thick accretion disk \citep{Grundstrom2007,Djura2008,Smith2011}. RY Scuti also exhibits smooth, periodic variability; however, its 11~d orbital period is much shorter than that of \nast, and its variability is not as consistent with a sine wave \citep{Djura2008}.

Given the significant contribution from the strong nebular emission lines to the observed photometry in \nast~(Table~\ref{tab:Prop}), it is possible that the photometric variability is linked to the obscuration of line-emitting regions from a circumbinary disk by \rev{a} circumstellar disk. Optical variability in this scenario could arise from a circumstellar disk obscuring a varying fraction of the emitting surface area \rev{of a} circumbinary disk, which requires the density distribution of the circumbinary disk to \rev{deviate from circular symmetry.} This may be expected if circumbinary material is associated with \nast's asymmetric [N\,{\sc ii}] nebula (Fig.~\ref{fig:Schem}). The observed IR variability may \rev{then} arise from disk eclipses of the observed circumstellar/circumbinary dust component around \nast~with an extent of $\lesssim150$ au \citep{Rajagopal2007}.\footnote{This has been updated with the recent $d=3.03$~kpc distance estimate to \nast~\citep{Rate2020}, and we note that \citet{Rajagopal2007} had adopted a distance of 2~kpc.} A similar alternative scenario is that a circumstellar disk may block a significant fraction of the ionizing and dust-heating radiation field from the \rev{stripped-envelope, mass-donor} star. If the surrounding nebulae of dust and ionized gas are asymmetric, this scenario could also lead to optical/IR variability throughout the orbit of the central binary.

\rev{Another disk-related scenario, where the mass gainer in \nast~has an optically-thick accretion-disk like RY~Scuti, is that the photometric variability may arise from the disk precession. The emission from the disk could be powered by reprocessed light from the hot, ionizing star in \nast, which would be analogous to the 162-d ``super-orbital'' period from the Galactic microquasar SS~433 (e.g.,~\citealt{Fabrika2004}).}
\rev{However, such a scenario should present complex photometric variability modulated by a shorter-timescale orbital period. Although a detailed analysis of this hypothesis is beyond the scope of this work, we consider this scenario unlikely given \nast's smooth and stable sinusoidal variability.}

\subsubsection{Colliding-wind Dust Formation}
Mid-IR and X-ray observations of \nast~indicate the presence of ongoing or recent dust formation that may be due to colliding winds \citep{Rajagopal2007,Mauerhan2015}. 
Dust formation via colliding winds is common in late-type carbon-rich WR (WC) binaries \citep{Williams1987} and is also believed to occur in $\eta$~Car \citep{Smith2010}. In this process, the strong wind from the WR star collides with the weaker wind from an OB-star companion which creates a dense shock front that cools and forms dust in the wake of the OB star \citep{Usov1991, Tuthill2008}.
An inclined dust-forming WR binary can exhibit smooth IR variability owing to viewing perspective variations toward the dust-forming region throughout the system's orbit \citep{Monnier1999,Hendrix2016}. The line-of-sight optical depth to the dust-forming regions can vary with viewing angle due to obscuration from optically thick regions of the colliding-wind shock cone.
These effects are evident in the colliding-wind dust-forming WR system WR~98a (WC8-9vd; \citealt{Williams1995}), which exhibits a ``rotating" dusty pinwheel nebula and \rev{roughly} sinusoidal IR variability that are both consistent with a $\sim565$~d period \citep{Monnier1999}. \rev{HD~36402, a WC4(+O?)+O8I system in the Large Magellanic Cloud \citep{Moffat1990}, also exhibits dust formation and sinusoidal IR variability with a 5.1~yr photometric period \citep{Williams2013,Williams2019}.}

\rev{Observations of} \nast~\rev{do} not \rev{present} an obvious dusty pinwheel nebula; however, \rev{a dusty pinwheel} may be difficult to resolve owing to its shorter orbital period and high inclination. 
\rev{It is therefore plausible} that the near-IR and mid-IR variability from \nast~may be due to a mechanism similar to that seen in WR~98a \rev{and HD~36402}.

In \rev{the colliding-wind dust-formation} scenario, the optical variability observed from \nast~could be linked to obscuration of a circumbinary nebula by optically thick regions of the colliding-wind shock cone and/or newly formed dust. 
\rev{As mentioned in Sec.~\ref{sec:Eline}, the variability amplitudes of the optical and IR light curves may therefore be associated with the fraction of the circumstellar/circumbinary emission over the stellar continuum emission in each filter band. The filter bands that exhibit that highest variability amplitude are notably the PGIR $J$ band, which should be dominated by circumstellar dust, and the ZTF $r$, KAIT $R$, and ATLAS $o$ bands, which cover the brightest nebular emission line He\,{\sc i} $\lambda6678$ as well as H$\alpha$ and [N\,{\sc ii}] $\lambda6583$. The other filter bands that show lower amplitude variability are less influenced by the nebular emission and may capture variability influenced by the stellar continuum emission.}

We note that it is unlikely that the observed variability of \nast~arises from variable or episodic dust formation since we would expect anti-correlated IR and optical light curves, where newly forming IR-luminous dust obscures the optical emission.

\subsubsection{Ellipsoidal Variability}
Given the \rev{interpretation of binary} mass transfer in \nast~\citep{Mauerhan2015}, the stars and/or a circumstellar disk may be tidally distorted and exhibit ellipsoidal variability as the system reveals different cross-sections throughout its nearly edge-on orbit (e.g., \citealt{Morris1993}). Such variations may naturally reproduce some of the wavelength-dependent amplitudes of the light curve owing to temperature gradients along the surface of the elongated star or disk. 

It is important to note that if the \nast~light curve was modulated by ellipsoidal variability \rev{throughout a binary orbit}, its orbital period would be twice that of the period derived from the sinusoidal model \rev{because} two consecutive brightness peaks would correspond to the two phases of quadrature in a full orbit. Tidally elongated stars in such a wide orbital configuration would therefore have to exhibit large radii. Assuming a \rev{$2\times310=620$~d} orbital period, a combined stellar mass range of 20--100~M$_\odot$, and a binary mass ratio of 0.25--4.0, the radius of the tidally distorted star would have to approach its Roche-lobe radius of 220--\rev{710}~R$_\odot$ \citep{Eggleton1983}. \rev{In this scenario, the} tidally distorted star \rev{should therefore} possess a radius consistent with that of a supergiant. 

\subsubsection{\rev{Preferred Interpretation and Unresolved Issues}}
\rev{Given the evidence for dust formation,} we favor the interpretation that \nast's periodic variability is linked to variable line-of-sight optical depth effects from colliding-wind dust formation. 
\rev{The longer timescale variability suggested by the past $J$-band light curve (Fig.~\ref{fig:PastVar}) may be a product of stellar variability and/or fluctuations in the optical/UV radiation heating the circumstellar dust.}
We cannot conclusively rule out a possible contribution \rev{to \nast's periodic photometric variability} from disk eclipses \rev{and/or ellipsoidal variability}. 
\rev{In Sec.~\ref{sec:nature} \&~\ref{sec:mass}, we will argue that \nast~does not host a supergiant, which suggests that ellipsoidal variability from a tidally distorted star is unlikely.}

It is \rev{interesting} to note that almost all known massive dust-forming colliding-wind binaries exhibit a C-rich chemistry and host a WC star \citep{Crowther2007}.
Future work on determining the composition of circumstellar dust around \nast~will be important for distinguishing the chemistry of \rev{the enshrouded} star(s).

\rev{Unfortunately, it is difficult to deduce a comprehensive interpretation of \nast's variability that satisfies all of the observational constraints. Most notably, there are still} unresolved issues on how to account for the apparent phase delay between different filters (Fig.~\ref{fig:Phase}, \textit{right}; Table~\ref{tab:Prop}) as well as \rev{the origin of longer timescale variability suggested by the historical $J$-band light curve} (Fig.~\ref{fig:PastVar}). 
\rev{We can speculate on the origin of the 17-day phase offset between the PGIR $J$ and ZTF $r$ light curves (Fig.~\ref{fig:Phase}, \textit{right}), where the phase delay timescale corresponds to a light-travel distance of 2940 AU, or 1\farcs0 at a distance of 3.03~kpc. Given the $\sim1''$ extent of bright circumstellar material in the central regions of \nast's nebula (Fig.~\ref{fig:Schem}, \textit{top left}), it is possible that the phase delay corresponds to the light-travel time between stellar and/or circumstellar components of \nast.}

\rev{Owing to} the complexity \rev{and obscured nature} of \nast, further observations and theoretical modeling are needed to resolve these issues and test our hypotheses. Three-dimensional radiative transfer models would be valuable for exploring how variations in line-of-sight optical depths from colliding-wind dust formation affect \rev{the} optical and IR light curves \rev{as well as the multi-epoch mid-IR BASS spectra}.
Continued multiwavelength photometric as well as spectroscopic monitoring of \nast~will also be important for resolving the origin(s) of its variability.

\subsection{On the Nature of \nast: An Ofpe/WN9 Eruption?}
\label{sec:nature}
\rev{Assuming that the photometric variability is modulated by the orbital motion of a binary in \nast, we can speculate on its nature in the context of the mass-transfer interpretation presented by \citet{Mauerhan2015}.}
The $\sim2$--4~au \rev{binary} orbital separation~(Sec.~\ref{sec:VarOrg}) allows us to estimate the Roche-lobe radius that the mass-donor star must have approached in order to trigger mass transfer. \rev{Adopting} a binary mass ratio of 0.25--4.0, the Roche-lobe radius is in the range 140--\rev{450}~R$_\odot$. Although such a large stellar radius \rev{might} suggest an RSG mass donor, 
\rev{an RSG is not consistent with the high temperatures ($\gtrsim30,000$ K) inferred from \nast's high-excitation nebular emission lines, and there are no spectroscopic RSG features in the the IR where its emission should be directly observable \citep{Crowther1999}.}

\citet{Crowther1999} suggested that the \rev{highly CNO-processed} composition of the nebula \rev{and the temperature of the hot ionizing source} is consistent with an \rev{early-type} nitrogen-rich WR (WN) star; however, they demonstrated that \nast~does not exhibit any broad features consistent with an \rev{early} WN star.
\rev{It is important to note that the absence of spectroscopic confirmation of WR features does not preclude the presence of a WR star in \nast. The LBV system $\eta$ Car is also enshrouded by dense circumstellar material and hosts an unseen hot massive companion that is thought be a WR star (see \citealt{Smith2018} and references therein).}

Interestingly, \citet{Smith2020} presented observations of an Ofpe/WN9 star, which are stars that show properties intermediate between those of Of and WN stars \citep{Walborn1977,Bohannan1989}, and confirmed that they can undergo an LBV-like eruption. During this eruptive phase, the radius of the Ofpe/WN9 star can expand to double the size of its quiescent radius \citep{Smith2020}.
Ofpe/WN9 stars in the Galactic center exhibit radii of $R_{2/3} \approx 35$--80~$R_\odot$, which corresponds to the radius where $\tau_\mathrm{Rosseland}=2/3$ \citep{Martins2007}. 
We therefore suggest that mass transfer in \nast~\rev{could} have been triggered during an eruptive phase from an Ofpe/WN9 star that is now obscured or outshined by dense circumbinary or circumstellar material. 
\rev{The massive RLOF binary RY Scuti is also believed to have undergone past LBV-like eruptions that led to the formation of its toroidal circumstellar nebula \citep{Smith2011}.}

Earlier, \citet{vdH1989,vdH1997} had proposed that \nast~may host an Ofpe/WN9 star; however, \rev{this was based only on the nebular spectrum.} \citet{Crowther1999} instead suggested that \nast~hosts an early-type WN star based on the high temperatures required for the ionization source of its nebula.
Our \rev{Ofpe/WN9} hypothesis on the nature of \nast~is still consistent with the interpretation by \citet{Crowther1999}:
after its eruption, the \rev{mass donor} may no longer resemble an Ofpe/WN9 star and could have transitioned to a hotter, earlier-type WN phase like the Ofpe/WN9 star MCA-1B studied by \citet{Smith2020}. It is also possible that envelope stripping \rev{may have led to} a more advanced chemical composition than the WN phase such as the WC phase. \rev{However,} the deficiency of carbon in the circumstellar environment around \nast~would require a very recent transition to a WC star.

\rev{We can obtain insight on the enshrouded star(s) in \nast~from radio observations taken by the \textit{Karl G.~Jansky} Very Large Array (VLA) telescope in the Global View of Star Formation in the Milky Way (GLOSTAR) survey \citep{Medina2019}. Interestingly, \nast~exhibits a radio spectral index of $\alpha=0.61\pm0.10$ between 4~Ghz and 8~GHz, consistent with the spectral index expected for free-free emission from an ionized stellar wind ($\alpha\approx0.6$; \citealt{Wright1975,Panagia1975}). Such thermal radio emission properties are commonly identified from the ionized winds of OB and WR stars \citep{Abbott1986,Bieging1989}. We can then estimate the mass-loss rate of the ionized wind given the theoretical relation between the observed radio flux density $F_\nu$ and the stellar wind parameters \citep{Wright1975}:}

\begin{equation}
    F_\nu=2.32\times10^4\left(\frac{\dot{M} Z}{v_\infty \mu}\right)^{4/3} \left(\frac{\gamma g_\nu \nu}{d^3}\right)^{2/3}\,\mathrm{mJy},
\label{eq:MLR1}
\end{equation}

\noindent
\rev{where $\dot{M}$ is the mass-loss rate in M$_\odot$ yr$^{-1}$, $v_\infty$ is the wind velocity in km s$^{-1}$, $\mu$ is the mean molecular weight, $Z$ is the mean ionic charge, $\gamma$ is the number of electrons per ion, $g_\nu$ is the Gaunt factor at observed frequency $\nu$, and $d$ is the distance in kpc.}
\rev{In order to estimate the mass-loss rate from the \nast~radio counterpart, we assume that $\mu \approx4.0$, which is consistent with H-poor winds, and $Z = 1.1$, $\gamma = 1.1$, and $g_{5.8}\approx5.0$ \citep{Leitherer1997}. Given the absence of any obvious stellar emission features from \nast's optical and near-IR spectrum \citep{Crowther1999}, we must provide a guess for the wind velocity associated with the radio counterpart. Since the wind is most likely ionized by a hot central WR or OB star, we adopt a wind velocity of $1000$ km s$^{-1}$. From \nast's observed 5.8~GHz flux density of $F_{5.8~{\rm GHz}}=4.38\pm0.23$ mJy \citep{Medina2019}, we then estimate a mass-loss rate of}

\begin{equation}
    \dot{M}\approx 2\times10^{-4}\left(\frac{v_\infty}{1000\,\mathrm{km}\, \mathrm{s}^{-1}}\right)\,\mathrm{M}_\odot\,\mathrm{yr}^{-1}.
\label{eq:MLR2}
\end{equation}

\noindent
\rev{This mass-loss rate estimate is likely uncertain by factors of a few given \nast's unknown wind velocity. However, we note that a mass-loss rate of $\sim10^{-4}$ M$_\odot$ yr$^{-1}$ is consistent with the upper range of mass-loss rates exhibited by Galactic WN stars \citep{Hamann2019}.}

\subsection{No \rev{(More)} Mass Transfer in \nast\rev{?}}
\label{sec:mass}

\rev{In this section, we address whether \nast~is undergoing RLOF mass transfer assuming the observed photometric variability is modulated by the orbital motion of the interacting (or previously interacting) binary. In order to assess the current state of mass transfer,} we estimate the radius of the \rev{hot, ionizing star in \nast} and compare it to the Roche-lobe radius ($R_L\approx140$--\rev{450}~R$_\odot$) derived above. 

First, we obtain an estimate of the \rev{stellar} luminosity from \nast's observed brightness $V=14.4$ mag. By adopting the interstellar \rev{and/or circumstellar} extinction correction toward \nast~by \citet{Crowther1999} of $A_V=6.5$ mag and using the {\it Gaia}-derived distance of $d=3.03$~kpc \citep{Rate2020}, the absolute magnitude of \nast~is $M_V=-4.5$. 
The 30,000~K lower limit of the temperature derived for the ionizing source in \nast~by \citet{Crowther1999} from optical spectroscopy \rev{implies a} bolometric correction is $BC_V = -3$~mag\rev{, which indicates a stellar luminosity of} $L \approx 7.9 \times 10^4$~L$_\odot$. The radius of such a 30,000~K star \rev{would be} $R_* \approx 10$ R$_\odot$, \rev{an order} of magnitude \rev{less than} the \rev{estimated} Roche-lobe radius. \rev{Stellar} temperatures higher than 30,000~K will require larger bolometric corrections and thus lead to higher \rev{stellar} luminosities\rev{; however,} the resulting radii will be smaller than that of \rev{the} 30,000~K star.
For example, a star with a temperature of 200,000~K, which is the upper limit of the ionizing source temperature derived by \citet{Crowther1999}, would imply a luminosity of $L \approx 3.1 \times 10^6$~L$_\odot$ and a stellar radius of $R_* \approx 1.5$~R$_\odot$ given the appropriate bolometric correction of $BC_V = -7$~mag. Based on these constraints, 
we \rev{suggest that \nast~is not currently undergoing binary mass transfer.}

\begin{deluxetable}{p{2.5cm}p{4.5cm}p{0.5cm}}
\tablecaption{\rev{\nast~Properties Summary}}
\tablewidth{0.98\linewidth}
\tablehead{Property & Value & Ref}
\startdata
Eq. Coordinates (J2000) & $18^{\rm h}52^{\rm m}17.55^{\rm s}$, $+00^\circ 59' 44\farcs3$ & 1\\
Gal.~Coordinates (J2000) & 033.9154$^\circ$, +00.2639$^\circ$ & 1\\
PM $\mu_\alpha\,\mathrm{cos}\,\delta$ & $-0.271\pm0.080$ mas yr$^{-1}$ & 1\\
PM $\mu_\delta$ & $-0.614\pm0.074$ mas yr$^{-1}$ & 1\\
Distance & $3.03^{+0.60}_{-0.45}$ kpc& 2\\
Spec. Classifications & WN10 \newline Ofpe/WN9, B[e], O[e] \newline Early WR + $\eta$ Car-like Nebula? & 3, 4\newline 5, 6 \newline 7\\
$A_V$ & 6.5 mag & 7\\
Var.~Period & $309.7\pm0.7$ d& 8 \\
$F_{5.8GHz}$ & $4.38\pm0.23$ mJy & 9 \\
Spec. Index, $\alpha$ & $0.61\pm0.10$ & 9 \\
$\dot{M}$ & $\sim10^{-4}$ M$_\odot$ yr$^{-1}$ & 8 \\
$L_X$ & $\sim10^{33}$ erg s$^{-1}$ & 10 \\ 
$L_\mathrm{bol}$ & $\sim10^5-10^6$ L$_\odot$ & 7, 8 \\
$B$ ($\Delta m_B$) & 16.208 (0.034) & 8\\
$g$ ($\Delta m_g$) & 15.349 (0.049) & 8\\
$c$ ($\Delta m_c$) & 14.370 (0.054) & 8\\
$V$ ($\Delta m_V$) & 14.414 (0.066) & 8\\
$R$ ($\Delta m_R$) & 12.850 (0.114) & 8\\
$r$ ($\Delta m_r$) & 12.775 (0.126) & 8\\
$o$ ($\Delta m_o$) & 12.608 (0.105) & 8\\
$I$ ($\Delta m_I$) & 12.303 (0.054) & 8\\
$J$ ($\Delta m_J$) & 9.632 (0.156) & 8\\
$J_\mathrm{Hist}$ ($\Delta m_{J\mathrm{,Hist}}$) & 9.25 (0.25) & 8\\
$H_\mathrm{Hist}$ ($\Delta m_{H\mathrm{,Hist}}$) & 8.18 (0.33) & 8\\
$K_\mathrm{Hist}$ ($\Delta m_{K\mathrm{,Hist}}$) & 6.17 (0.29) & 8\\
$L'_\mathrm{Hist}$ ($\Delta m_{L'\mathrm{,Hist}}$) & 3.87 (0.18) & 8\\
\enddata
\tablecomments{\rev{References: 1 - \citet{Gaia18}, 2 - \citet{Rate2020}, 3 -- \citet{Nassau1963}, 4 -- \citet{Massey1983}, 5 -- \citet{vdH1989}, 6 -- \citet{vdH1997}, 7 -- \citet{Crowther1999}, 8 -- This work, 9 -- \citet{Medina2019}, 10 -- \citet{Mauerhan2015} }}
\label{tab:summary}
\end{deluxetable}

\section{Conclusions}
\label{sec:conclusions}
\rev{A comprehensive summary of \nast's properties including the results from our work are provided in Table~\ref{tab:summary}. In this paper,} we presented the discovery of a \rev{$310$~d photometric variability} period from \nast~with optical and near-IR light curves obtained between 2014 July and \rev{2021 July} (Fig.~\ref{fig:LC}).
\rev{The best-constrained period, $P=309.7\pm0.7$~d, is derived from a sinusoidal fit to the ZTF $r$-band light curve.}
The \rev{amplitude of \nast's} sinusoidal variability \rev{differed at different} wavelengths\rev{, and we identify a} significant phase shift between the sine models fit to the $r$-band and $J$-band light curves (Table~\ref{tab:Prop}; Fig.~\ref{fig:Phase}\rev{, \textit{right}}).
Based on previously cataloged optical emission lines from the nebula surrounding \nast~\citep{Crowther1999}, we determined that nebular emission lines contribute $\sim5$--30\% to the \rev{optical} photometry (Table~\ref{tab:Prop})\rev{. The fraction of emission originating from circumstellar material} may therefore be linked to the differing variability amplitudes \rev{observed across the filter} wavelengths. 
Historical IR light curves over 1983--1989 \rev{show that the past variability was consistent with the present-day period of $310$~d. However, the past $J$-band brightness and variability amplitudes (Tab.~\ref{tab:HistLCFit}) were greater than that measured by the recent PGIR light curves, suggesting that \nast\ exhibits variability on longer ($\gtrsim$ decade) timescales.}
Mid-IR \rev{AEOS/BASS} spectra of \nast, which were taken semicontemporaneously with ZTF\rev{, ATLAS,} and PGIR observations, demonstrated that its 3--13~$\mu$m emission is also variable and appears to be correlated with the optical/near-IR variability (Fig.~\ref{fig:SED}).

\rev{We discussed possible single-star origins for \nast's photometric variability. Stellar pulsations are an unlikely origin given the inconsistency between the observed and expected trend in the variability amplitude as a function of wavelength (see Fig.~\ref{fig:Amp}). We also considered rotational modulation in a rare class of highly magnetized O-stars (Of?p) as a possible origin of single-star variability. However, the amplitude of photometric variations from Of?p stars should show no significant deviations as a function of wavelength, which is also discrepant from the observations of \nast. We therefore postulate that the observed $310$~d} optical/near-IR periodic, sinusoidal variability is associated with the orbital period of a binary system in \nast.
Although the properties of the stars in this binary system are still unknown, we use the \rev{$310$~d} orbital period to estimate an orbital separation of $\sim2$--4~au assuming \rev{a combined binary mass ranging between} 20--100~M$_\odot$ (Fig.~\ref{fig:Schem}, top).

Given the nearly edge-on geometry of \nast's extended nebula \citep{Mauerhan2015}, we present possible \rev{binary} origins of variability including (1) eclipses of a circumbinary disk from an optically and geometrically thick circumstellar disk around one of the stars, (2) variations in the line-of-sight optical depth to optical/near-IR emitting regions due to colliding-wind dust formation, and (3) ellipsoidal variability from a tidally elongated star (Fig.~\ref{fig:Schem}, bottom). We favor the colliding-wind dust formation hypothesis; however, future work with 3D radiative transfer modeling is crucial to reproduce optical and IR light curves that incorporate the line-of-sight optical depth variability throughout the orbit of a highly inclined dust-forming colliding-wind binary.
\rev{Ultimately, the origin(s) of \nast's photometric variability is still an open question. Further work and more observations are needed to develop a comprehensive interpretation that satisfies all of the observational constraints including the phase offsets and historical IR variability.} 

\rev{Assuming that \nast's periodic photometric variability reflects a $310$~d binary orbital period, we addressed whether the binary is undergoing RLOF mass transfer.} Based on the CNO-enriched material in \nast's nebula and the estimated range of sizes for the Roche-lobe radius of the mass donor star in \nast\ ($\sim140$--\rev{450}~R$_\odot$), we suggest that mass transfer \rev{could have been} triggered by an LBV-like eruptive phase in an Ofpe/WN9 star in \nast~that filled its Roche lobe as it expanded. This interpretation is bolstered with recent observations by \citet{Smith2020} that confirmed Ofpe/WN9 stars can undergo LBV-like eruptions and expand to double the size of their quiescent radius. Owing to the mass loss from its outer envelope, it is possible that the \rev{mass donor} star in \nast~no longer resembles an Ofpe/WN9 star. The \rev{donor} star may have transitioned to either a hotter, earlier WN spectral subtype or a more chemically advanced WC phase. 
\rev{Radio observations of \nast~in the GLOSTAR survey \citep{Medina2019} reveal a spectral index consistent with free-free emission from an ionized stellar wind and reinforces the interpretation that \nast~hosts a hot massive star. Given the relation between the observed radio flux and stellar wind parameters provided by \citet{Wright1975}, we estimated a mass-loss rate of $\sim10^{-4}$ M$_\odot$ yr$^{-1}$,consistent with the upper range of mass-loss rates exhibited by Galactic WN stars. }

Based on the inferred orbital separation \rev{and the} radius estimates of the star that dominates the \rev{ionizing flux and} $V$-band emission, \rev{we suggest that such a binary system is no longer undergoing mass transfer.}
\rev{\nast~still remains an enigmatic system that could provide new insights into an important transitional evolutionary phase of massive stars}. Multiwavelength follow-up observations and continued spectroscopic and photometric monitoring will therefore be valuable for addressing the open questions on the nature of \nast.

\acknowledgments
We thank T.~Jayasinghe for discussion of the technical details of the filter properties in the ASAS-SN survey.
\rev{We also thank M.~Munoz for an enlightening discussion on Of?p stars.}
R.M.L. acknowledges the Japan Aerospace Exploration Agency's International Top Young Fellowship (ITYF).
A.F.J.M. is grateful for financial assistance from NSERC (Canada).
M.M.K. acknowledges the Heising-Simons foundation for support via a Scialog fellowship of the Research Corporation.
M.M.K. and A.M.M. acknowledge the Mt. Cuba Astronomical Foundation. 
A.V.F. is grateful for financial assistance from the TABASGO Foundation, the Christopher R. Redlich Fund, the U.C. Berkeley Miller Institute for Basic Research in Science (in which he is a Miller Senior Fellow), and many individual donors.
M.M.K. acknowledges generous support from the David and Lucille Packard Foundation. 
J.S. is supported by an Australian Government Research Training Program (RTP) Scholarship.

Palomar Gattini-IR (PGIR) is generously funded by Caltech, Australian National University, the Mt. Cuba Astronomical Foundation, the Heising-Simons Foundation, and the Binational Science Foundation. PGIR is a collaborative project among Caltech, Australian National University, University of New South Wales, Columbia University, and the Weizmann Institute of Science. 
Based in part on observations obtained with the Samuel Oschin 48-inch Telescope at the Palomar Observatory as part of the Zwicky Transient Facility project. ZTF is supported by the National Science Foundation (NSF) under grant AST-1440341 and a collaboration including Caltech, IPAC, the Weizmann Institute for Science, the Oskar Klein Center at Stockholm University, the University of Maryland, the University of Washington, Deutsches Elektronen-Synchrotron and Humboldt University, Los Alamos National Laboratories, the TANGO Consortium of Taiwan, the University of Wisconsin at Milwaukee, and Lawrence Berkeley National Laboratories. Operations are conducted by Caltech Optical Observatories, IPAC, and the University of Washington.

We thank the Las Cumbres Observatory and its staff for its continuing support of the ASAS-SN project. LCOGT observations were performed as part of DDT award 2019B003 to E.G. ASAS-SN is supported by the Gordon and Betty Moore Foundation through grant GBMF5490 to the Ohio State University, and NSF grants AST-1515927 and AST-1908570. Development of ASAS-SN has been supported by NSF grant AST-0908816, the Mt. Cuba Astronomical Foundation, the Center for Cosmology and AstroParticle Physics at the Ohio State University, the Chinese Academy of Sciences South America Center for Astronomy (CAS- SACA), the Villum Foundation, and George Skestos.

UKIRT is owned by the University of Hawaii (UH) and operated by the UH Institute for Astronomy; operations are enabled through the cooperation of the East Asian Observatory. When the data reported here were acquired, UKIRT was operated by the Joint Astronomy Centre on behalf of the Science and Technology Facilities Council of the U.K.

Research at Lick Observatory is partially supported by a generous gift from Google. KAIT and its ongoing operation were made possible by donations from Sun Microsystems, Inc., the Hewlett-Packard Company, AutoScope Corporation, Lick Observatory, the NSF, the University of California, the Sylvia \& Jim Katzman Foundation, and the TABASGO Foundation. 

\rev{This work has made use of data from the Asteroid Terrestrial-impact Last Alert System (ATLAS) project. The Asteroid Terrestrial-impact Last Alert System (ATLAS) project is primarily funded to search for near-Earth objects (NEOs) through NASA grants NN12AR55G, 80NSSC18K0284, and 80NSSC18K1575; byproducts of the NEO search include images and catalogs from the survey area. This work was partially funded by Kepler/K2 grant J1944/80NSSC19K0112 and HST GO-15889, and STFC grants ST/T000198/1 and ST/S006109/1. The ATLAS science products have been made possible through the contributions of the University of Hawaii Institute for Astronomy, the Queen’s University Belfast, the Space Telescope Science Institute, the South African Astronomical Observatory, and The Millennium Institute of Astrophysics (MAS), Chile.}

This research made use of Astropy,\footnote{http://www.astropy.org} a community-developed core Python package for Astronomy \citep{Astropy2013, Astropy2018}.

%

\vspace{5mm}
\facilities{PGIR, \rev{ATLAS}, KAIT, ZTF, ASAS-SN, AEOS (BASS), UKIRT, ESO 1-m}

\end{document}